\documentclass{aa}
\usepackage{etoolbox}

\usepackage{graphicx}
\usepackage{txfonts}

\usepackage{float}
\usepackage{pdflscape} 
\usepackage{afterpage} 

\usepackage{tabularx}

\usepackage{geometry}
\usepackage{caption}
\usepackage{lipsum}

\makeatletter
\renewcommand*\aa@pageof{, page \thepage{} of \pageref*{LastPage}}
\usepackage{multirow}
\usepackage{lscape}
\usepackage{orcidlink}

\newcommand*{\bba}{$^{\scriptstyle 3\mathrm{D}}$Barolo\xspace}

\newcommand*{\hi}{H\,{\sc i}\xspace}

\usepackage{hyperref}
\hypersetup{
     colorlinks = true,
     linkcolor = blue,
     anchorcolor = blue,
     citecolor = blue,
     filecolor = blue,
     urlcolor = blue
     }
\begin{document} 

   \title{
   \textit{FRIDA}: Finding Rotation In ultra-Diffuse gAlaxies} 
   \subtitle{I. Programme overview, resolved gas density profiles, and \hi mass-size relation}
   \titlerunning{\textit{FRIDA} I. \hi density profiles and mass-size relation of gas-rich UDGs}
\authorrunning{Mancera Pi\~na et al.}

   \author{Pavel E. Mancera Pi\~na\inst{1,2}\fnmsep\thanks{\email{Pavel.ManceraPina@ugent.be}}\orcidlink{0000-0001-5175-939X}, 
Tom Oosterloo\inst{3}\orcidlink{0000-0002-0616-6971}, 
Filippo Fraternali\inst{4}\orcidlink{0000-0002-0447-3230}, 
Elizabeth A. K. Adams\inst{3,4,5}\orcidlink{0000-0002-9798-5111},
Andrea Afruni \inst{6,7}\orcidlink{0000-0002-2858-6950}, 
Cecilia Bacchini \inst{8}\orcidlink{0000-0002-8372-3428}, 
Enrico M. Di Teodoro\inst{6,7}\orcidlink{0000-0003-4019-0673},
Lexi Gault\inst{9}\orcidlink{0000-0002-2492-7973},
Anastasia A. Ponomareva\inst{10,11}\orcidlink{0000-0003-4100-0173},\\ 
Justin I. Read\inst{12}\orcidlink{0000-0002-1164-9302}, \and
Barbara Šiljeg\inst{3,4}\orcidlink{0000-0001-5391-8286}.
\\
          }
   \institute{Leiden Observatory, Leiden University, P.O. Box 9513, 2300 RA, Leiden, The Netherlands 
   \and Department of Physics and Astronomy, Ghent University, Proeftuinstraat 86 N3, B-9000 Ghent, Belgium
   \and ASTRON, Netherlands Institute for Radio Astronomy, Oude Hoogeveensedijk 4, 7991 PD Dwingeloo, the Netherlands
    \and Kapteyn Astronomical Institute, University of Groningen, Postbus 800, 9700 AV Groningen, The Netherlands
      \and SKA Observatory, SKA-Mid Science Operation Centre, 2 Fir St, Observatory, Cape Town 7925, South Africa
    \and Dipartimento di Fisica e Astronomia, Università di Firenze, Via G. Sansone 1, 50019 Sesto Fiorentino, Firenze, Italy
   \and INAF – Osservatorio Astrofisico di Arcetri, Largo E. Fermi 5, Firenze 50125, Italy
   \and DARK, Niels Bohr Institute, University of Copenhagen, Jagtvej 155, 2200 Copenhagen, Denmark
          \and Department of Astronomy, Indiana University, 727 East 3rd Street, Bloomington, IN 47405, USA
   \and University of Hertfordshire, College Ln, Hatfield AL10 9AB, United Kingdom
   \and Oxford Astrophysics, Denys Wilkinson Building, University of Oxford, Keble Road, Oxford OX1 3RH, United Kingdom
   \and Department of Physics, University of Surrey, Guildford, Surrey GU2 7XH, United Kingdom
   }

   \date{}

  \abstract{
Gas-rich ultra-diffuse galaxies (UDGs) provide powerful and novel tests of galaxy formation models at low stellar surface densities. However, the structure of their neutral atomic hydrogen (\hi) reservoirs remains poorly constrained, despite \hi being their dominant baryonic component, hampering our understanding of this population.

We present \textit{FRIDA} (Finding Rotation In ultra-Diffuse gAlaxies), a 250-hour MeerKAT programme complemented by existing Very Large Array data, providing high-resolution \hi imaging of 12 gas-rich UDGs. By expanding the number of UDGs with high-resolution \hi observations from two to 12, \textit{FRIDA} provides a foundation for detailed studies of the gas structure, evolution, and dark matter content of diffuse gas-rich dwarfs. In this paper, we introduce our sample and delve into the associated interferometric observations and data reduction. We derive \hi fluxes, \hi masses ($M_{\rm HI}$), and, for the first time for a large UDG sample, well-resolved \hi surface density profiles ($\Sigma_{\rm HI}(R)$). We find extended (${\sim}7{-}30\,\rm{kpc}$ in radius) and regular \hi emission in our targets, and we report the discovery of a satellite around one of them. Most UDGs in \textit{FRIDA} have $\Sigma_{\rm HI}(R)$ profiles with flat or shallow inner parts reaching $\Sigma_{\rm HI}(R)\approx4\,M_\odot\,\rm{pc^{-2}}$ near their centres, which then decline exponentially in their outer parts. By modelling the \hi morphology of the UDGs, we find a median inclination of $43^\circ$ for the sample, in agreement with theoretical expectations for disc galaxies under UDG selection effects. Moreover, we place our UDGs in the $M_{\rm HI}{-}R_{\rm HI}$ and $M_{\rm HI}{-}R_{\rm 50,HI}$ mass-size relations defined by reference literature samples, where $R_{\rm HI}$ is the isophotal radius at which $\Sigma_{\rm HI}(R)\,{=}\,1\,M_\odot\,\rm{pc^{-2}}$, and $R_{\rm 50,HI}$ is the radius enclosing $50\%$ of $M_{\rm HI}$. While our UDGs lie within the scatter of the relations, they systematically fall on their upper side, being slightly (by ${\sim}\,15{-}25\%$) more extended and of lower mean $\Sigma_{\rm HI}$ (by ${\sim}\,25{-}35\%$) than average galaxies at similar $M_{\rm HI}$. 
  Moreover, we find that $\Sigma_{\rm HI}(R/R_{\rm HI})$ and $M_{\rm HI}({<}\,R)/M_{\rm HI}$ for UDGs follow an intermediate trend between late-type discs and early-type galaxies, with a large overlap with Sa galaxies.
  Overall, our observations provide novel \hi data for UDGs and impose important constraints on models and simulations that produce UDG-like systems.
}


   \maketitle 

\section{Introduction}
\label{sec:intro}
Ultra-diffuse galaxies (UDGs, \citealt{vandokkum2015}) are low-surface-brightness (LSB) galaxies with extended light distributions. The precise definition to classify galaxies as UDGs varies in the literature, but the most used convention is to have an optical effective radius $R_{\rm e}\,{\geq}\,1.5\,\mathrm{kpc}$ and a (central or mean effective) surface brightness fainter than $\mu\,{\sim}\,24\,\mathrm{mag\, arcsec^{-2}}$ in the $g$ or $r$ bands (e.g. \citealt{vandokkum2015,vanderburg2016}). UDGs are present in all galactic environments, from isolated cases to those inhabiting massive galaxy clusters (e.g. \citealt{koda2015,roman_udslargescale,roman_outsideclusters,jones_huds,udgs_kiwics,roman_udginvoid, smudges,iodice_lewis,makda2025,marleau2025}). While cluster UDGs are gas-poor, field UDGs host significant reservoirs of cold gas, in particular \hi, neutral atomic hydrogen \citep{leisman2017,janowiecki2019,karunakaran2024}. The relationship between the gas-poor and gas-rich UDG populations remains unclear; it is not known whether they represent different evolutionary stages of the same class of objects or distinct populations altogether (e.g. \citealt{grishin2021,buzzo2025,taylor2026} and references therein).

\looseness=-1
While large LSB galaxies have been known for decades (e.g., \citealt{sandage1984,binggeli1995,conselice2018}), interest in UDG-like galaxies has grown significantly over the last decade, as their extreme properties are reproduced in modern models and simulations using very different modelling prescriptions. For instance, it has been argued that UDGs form due to either particularly bursty star formation histories, overly-efficient globular cluster formation, a high-spin angular momentum dark matter halo, and/or peculiar merger or accretion histories (e.g. \citealt{amorisco2016,nihao_udgs,fire_udgs,jiang_udgs,silk_udgs,nihao_udgs_kinematics,wright_udgs,trujillogomez_udgs,demao,moreno2022,vandokkum_bullet,vannest2022,benavides2023,kravtsov2024,udgs_eagle}). Being UDGs broadly defined in terms of the size-surface brightness parameter space, the observed UDG population is likely a mixed bag of galaxies with diverse formation mechanisms (e.g. \citealt{vannest2022,benavides2023,buzzo2025}).

Over the last few years, a handful of gas-rich UDGs have been reported to exhibit puzzling dynamical properties\footnote{For puzzling gas-poor UDGs see e.g. \cite{vandokkum_DF2,montes2024,buzzo2026}, and references therein.}: Resolved (but still low-resolution) \hi kinematic measurements show that some gas-rich UDGs have disc-like rotation \citep{huds2019,sengupta2019,huds2020,lexi,barbara_udgs}, but the overall rotational speeds of most of them are low for their baryonic mass, which makes many of them outliers of the baryonic Tully-Fisher relation (BTFR, e.g. \citealt{mcgaugh2000,anastasia_SED}), connecting the baryonic mass of galaxies with their outer circular speeds.

Obtaining high-resolution observations of gas-rich UDGs is very challenging due to their low surface densities and relatively large distances (${\sim}\,30{-}100\,\rm{Mpc}$). However, such observations may be fundamental: The only two gas-rich UDGs (AGC~114905 and AGC~242019) with high-resolution (more than eight independent beams across their diameters) \hi observations have rotation curves suggesting extremely low dark matter concentration parameters (\citealt{shi2021,agc114905}, see also \citealt{sengupta2019} for an additional case at intermediate resolution) that seem in tension with the cornerstone halo concentration–mass relation expected in the cold dark matter (CDM) model. While the inner dark matter densities of classical dwarf galaxies are broadly consistent with hydrodynamical CDM simulations including stellar feedback \citep{mancera2026b}, the unusually low concentrations inferred for these UDGs may be difficult to explain within the same framework, potentially favouring alternative dark matter models \citep{demao,nadler2023,agc114905_deep}. Additionally, the discs of UDGs are powerful tools for testing the limit conditions under which galaxy discs can remain locally and globally stable \citep{sellwood_agc114905,ceci_q3d,afruni2025} and undergo star formation \citep{kado-fong_2022_1,kado-fong_2022_2}. Given all of the above, gas-rich UDGs represent an extremely promising avenue for testing dark matter models and our theories of galaxy formation and evolution. However, with only two gas-rich UDGs having high-resolution \hi data, it remains impossible to fully exploit this galaxy population.

To alleviate this limitation, we have carried out the programme \textit{FRIDA} (Finding Rotation In ultra-Diffuse gAlaxies), which we present here for the first time.
\textit{FRIDA} has exploited the MeerKAT and Karl G. Jansky Very Large Array (VLA) interferometers to map \hi in a sample of 12 gas-rich UDGs at high spatial and spectral resolution. \textit{FRIDA}'s \hi observations are complemented by ultra-deep optical imaging carried out with the 10.4--m Gran Telescopio Canarias (GTC), with a depth and quality that surpass those of typical optical observations of UDGs (see for instance \citealt{agc114905_deep}). \textit{FRIDA} has a broad scope, and science cases being developed include \emph{i)} modelling of the UDGs' gas kinematics to assess their position in fundamental scaling relations and constrain their formation mechanisms, \emph{ii)} dynamical modelling to assess the performance of different dark matter models, \emph{iii)} testing the limits of disc stability with simulations and state-of-the-art diagnostics, \emph{iv)} study the conditions of the interstellar medium at very low volume density and star formation surface density regimes, and \emph{v)} understanding the truncation, size, and spiral and clumpy features in their optical morphologies.

In this first paper, we introduce the galaxy sample that constitutes \textit{FRIDA} and present the first scientific results concerning the \hi properties of our UDGs. The rest of this paper is organised as follows. In Section~\ref{sec:data}, we provide an overview of \textit{FRIDA}, explain the sample selection and \hi observational campaign, and detail the data reduction process. In Section~\ref{sec:HIprop}, we derive the main \hi properties of the sample, including \hi masses and resolved density profiles. In Section~\ref{sec:mainresults}, we present and discuss our main scientific results. In particular, we delve into the observed distribution of inclination angles in UDGs, study the UDGs' \hi mass-size relations, and highlight the similarities and differences of their \hi distribution compared to other gas-rich galaxies.
Finally, in Section~\ref{sec:conclusions} we summarise our main findings. Throughout this work, we adopt a $\Lambda$CDM cosmology with $\Omega_{\mathrm{m},0} = 0.3$, $\Omega_{\Lambda,0} = 0.7$, and $H_0 = 70~\mathrm{km\,s^{-1}\,Mpc^{-1}}$.

\section{\textit{FRIDA}}
\label{sec:data}
\subsection{Overview and sample selection}

\textit{FRIDA} comprises deep, high-resolution \hi observations of 12 isolated gas-rich UDGs obtained with the MeerKAT and VLA interferometers. Moreover, \textit{FRIDA} UDGs are followed up by ultra-deep optical imaging with the GTC. The GTC data reach typical surface brightness depths of around $32\,\mathrm{mag\ arcsec^{-2}}$ in the $r$-band when measured at a $3\,\sigma$ level in boxes of $10\,\mathrm{arcsec}\,{\times}\,10\,\mathrm{arcsec}$ \citep{agc114905_deep}. The optical data set will be presented in an upcoming paper.

The UDGs in \textit{FRIDA} are drawn from the parent samples by \citet[][10 targets]{leisman2017} and \citet[][two targets]{janowiecki2019}, who created catalogues of gas-rich UDGs based on SDSS-DR12 \citep{sdss_dr12} and ALFALFA \citep{alfalfa40, alfalfa} optical and \hi data. Their UDGs were selected to have \hi detections, optical $R_{\rm e}\,{\geq}\,1.5\,\mathrm{kpc}$ and a mean effective surface brightness within $R_{\rm e}$ in the $r$-band $\langle \mu(R_{\rm e})\rangle_{r}\, {\geq}\, 24\,\mathrm{mag\,arcsec{^{-2}}}$; a minimum Hubble flow distance of $25\,\mathrm{Mpc}$ was imposed to reduce the impact of peculiar velocities.
The first step towards building \textit{FRIDA} consisted of exploiting existing VLA observations. Two UDGs have been observed with the VLA at relatively high spatial resolution (${>}\,8$ resolution elements across their discs' major axis): AGC~114905 \citep{agc114905} and AGC~242019 \citep{shi2021}.

\looseness=-1
However, what made it possible to enlarge the sample with high-quality data was the advent of MeerKAT and its superb sensitivity, which we exploited over 250 hours, as detailed below. To build our target sample for MeerKAT, we first selected galaxies from the catalogues of \cite{leisman2017} and \cite{janowiecki2019} with declinations $\delta\,{<}\,5^\circ$, for which MeerKAT has a better $uv$-coverage and point spread function. Moreover, we retained only those galaxies whose expected \hi angular diameters could be resolved by more than 10 MeerKAT beams; for this, we used the ALFALFA \hi masses and assumed \hi physical diameters from the calibration by \cite{wang16}, which were then converted to angular diameters. Finally, we selected galaxies spanning a range of apparent offsets from the BTFR to include galaxies with potentially different dynamical properties. 
For these BTFR offsets, we used ALFALFA for linewidths and \hi masses, and DECaLs imaging \citep{dey2019} for stellar masses and preliminary inclinations (using isophotal fitting as detailed in \citealt{barbara_udgs}). Although these measurements carry substantial uncertainties and caveats given the use of unresolved kinematics and DECaLs imaging (for UDGs, these data may not be deep enough to trace the faint underlying discs; see \citealt{agc114905_deep}), we used them as practical proxies to try to identify UDGs lying within the scatter of, moderately offset from, and significantly offset from the BTFR, to explore the physical processes underlying these deviations. 

We selected a number of galaxies that met the above conditions, and 10 of them were observed with MeerKAT, bringing the final sample size of \textit{FRIDA} to 12 gas-rich UDGs, as listed in Table~\ref{tab:beams}. While our sample is complete for now, any additional UDG we observe at high resolution in the future will be added to \textit{FRIDA}'s website\footnote{\url{https://pavel-mancera-pina.github.io/frida}}. As discussed in Appendix~\ref{app:neighbours}, our UDGs inhabit very low-density environments and have no massive nearby companions. 

\begin{figure}
    \centering
    \includegraphics[width=1\linewidth]{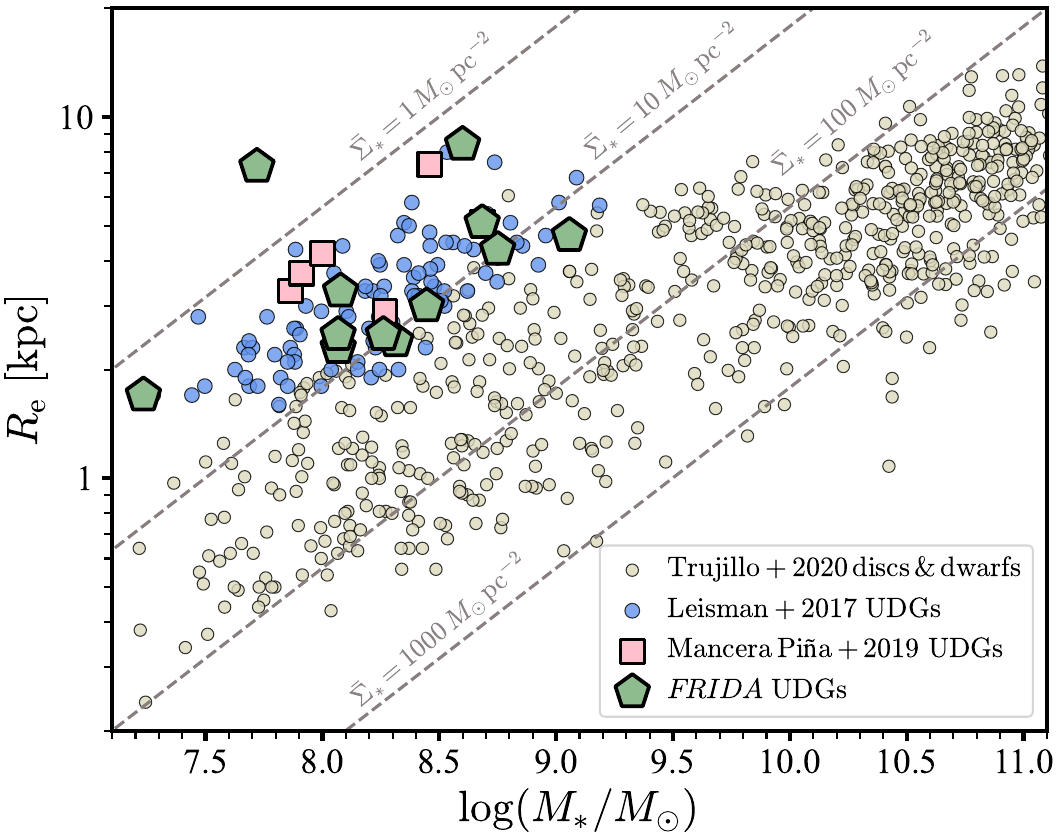}
    \caption{Stellar mass-size relation for \textit{FRIDA} UDGs compared to other galaxy samples. UDGs stand out as the dwarf galaxies with the largest $R_{\rm e}$ and the lowest average stellar surface densities within $R_{\rm e}$ (grey dashed lines show indicative values). The measurements of the UDGs in this plot are preliminary, as they are based on relatively shallow imaging and will be refined in our upcoming studies using GTC. 
    }
    \label{fig:reffmstar}
\end{figure}

To emphasise the low density of our UDG sample, in Figure~\ref{fig:reffmstar} we show the optical $M_\ast\,{-}\, R_{\rm e}$ relation for \textit{FRIDA} compared to the catalogue (dwarfs and discs only) from \cite{trujillo2020}, the UDG parent sample from \cite{leisman2017}, and the gas-rich UDGs studied in \cite{huds2019}, one of which (AGC~114905) is included in \textit{FRIDA}. The measurements for all UDGs are taken from \cite{leisman2017} and \cite{janowiecki2019} and are preliminary, since they were derived from SDSS imaging and will be refined with our upcoming GTC imaging. Still, they are informative to show the large optical sizes and low average stellar surface densities within $R_{\rm e}$ (dashed lines) of gas-rich UDGs relative to the general dwarf population. We note that since gas-poor and gas-rich UDGs are selected only because of their optical properties, both groups overlap in the $M_\ast\,{-}\, R_{\rm e}$ relation.  The figure also illustrates that \textit{FRIDA} UDGs bridge the gap in the stellar mass-size relation between the classical dwarf population and the most extreme UDGs studied in \cite{huds2019}. We show DECaLS RGB stamps for our galaxy sample in Appendix~\ref{app:optical}, although we will study the optical properties of our UDGs once we complete the analysis of our ultra-deep GTC imaging.

\subsection{\hi observations and data reduction}
In this section, we describe the observing strategy and data reduction of our interferometric data. Table~\ref{tab:beams} lists our galaxy sample, the beam properties for our \hi data cubes, and their rms noise per channel, all as described below.

\subsubsection{VLA data}
Two galaxies were observed with the VLA: AGC~114905 and AGC~242019. The data used for AGC~114905 and its reduction are fully described in \cite{agc114905}. Similarly, the details on the observations and data reduction of AGC~242019 are presented in \cite{shi2021}. 
For AGC~114905, we use the data cube from \cite{agc114905} and smooth it to a circular beam of 9 arcsec, thereby increasing the signal-to-noise ratio (S/N). The spectral resolution of the cube is $3.4\,\mathrm{km\,s^{-1}}$, and it has been Hanning-smoothed, so the channel width is half that. For AGC~242019, we use the data cube kindly provided by \cite{shi2021}, which we downsample to a pixel scale so that $b_{\mathrm{min}}\,{\approx}\,5$ pixels (instead of the original $b_{\rm min}\,{\approx}\,10$ pixels, with $b_{\mathrm{min}}$ the FWHM of the beam minor axis), and smooth it to a circular beam of 12 arcsec. The spectral resolution of this cube is $7\,\rm{km\,s^{-1}}$. The cube properties of both AGC~114905 and AGC~242019 are listed in Table~\ref{tab:beams}.

\subsubsection{MeerKAT data}
Our MeerKAT observations were obtained through a Pilot (ID SCI-20230907-PM-01; P.I. Mancera Piña) and a Large Programme (SCI-20241101-PM-01; P.I. Mancera Piña). Each galaxy was observed for 20 hours (25 hours including calibrators). In practice, the observations were split into five observing blocks. Each of these blocks consisted of the following steps: bandpass calibrator (10 minutes), four cycles of gain calibrator (2 minutes) + target (30 minutes), gain calibrator (2 minutes), bandpass calibrator (10 minutes), four cycles of gain calibrator (2 minutes) + target (30 minutes), gain calibrator (2 minutes), and bandpass calibrator (10 minutes). We used the c856M4k\_n107M correlator centred in the \hi line at the redshift of our targets, spanning a 107 MHz bandwidth with 32768 frequency channels spaced by $3.265\, \mathrm{kHz}$ (${\sim}\, 0.7\,\mathrm{km\,s^{-1}}$). This configuration was planned (and validated through our pilot programme) to ensure reaching an \hi surface density of $0.5\, M_\odot\,\mathrm{pc^{-2}}$ at a signal-to-noise larger than 3 at high spatial (${\sim}8\,\mathrm{arcsec}$) and spectral (${\sim}3\,\mathrm{km\,s^{-1}}$) resolution.

We downloaded the calibrated data from the MeerKAT SARAO archive and applied a channel binning of 2 for a (temporary) spectral resolution of $6.53\, \mathrm{kHz}$ (${\sim}1.4\,\mathrm{km\,s^{-1}}$). After corroborating their high quality, we adopted the default MeerKAT delay, flux, bandpass, gain, and phase calibrations (see \href{https://skaafrica.atlassian.net/wiki/spaces/ESDKB/overview}{The MeerKAT Knowledge Base}). The next steps for the data reduction employed a multi-stage pipeline within the \textsl{CASA} environment \citep{casa}. Specifically, we did as follows.
\begin{enumerate}
    \item We flagged poor visibilities to remove RFI artefacts using the task \textsl{flagdata}. For each of our five observing blocks, we did this in three steps: first, using the `quack' mode with a quack interval of 5 sec; second, using the `rflag' mode; and third, using the `extend' mode (the last two tasks using the default \textsl{CASA} parameters).
    
    \item Each observing block was regridded to a common barycentric reference frame (LSRK) using \textsl{mstransform}. This accounts for Earth's orbital motion and ensures that the emission from the target remains at fixed channel indices across all epochs, enabling a consistent treatment of the \hi line.

    \item We performed continuum subtraction in the visibility domain for each block individually with the task \textsl{uvcontsub}. We fitted line-free channels using a polynomial of order 0 or 1. Preference was given to 0th-order polynomials unless continuum residuals remained. 
    Following continuum subtraction, we recalculated the visibility weights for each block using \textsl{statwt}, restricted to the line-free frequencies. In this way, each observation was weighted by its actual noise variance, thereby optimising the sensitivity of the final combined dataset. 
    
    \item The calibrated, subtracted, and weighted visibilities were then jointly imaged using \textsl{tclean}. We cleaned the channels down to ${\sim}0.5\,\sigma$ using the `auto-multithresh' mask,
    and visually checked every major cycle to ensure that no oversubtraction occurred. We adopted a pixel scale such that $b_{\rm min}$ was sampled with ${\sim}5$ pixels. We restored to the common beam, and produced cubes with different robust Briggs weighting parameters $r$. After experimenting, we adopted fiducial cubes with $r\,{=}\,0.5$, which strike a fine balance between spatial resolution and sensitivity and will be used throughout this paper and in our future kinematic analysis. If continuum artefacts remained in the cube, we used \textsl{imcontsub} to perform continuum subtraction in the image plane. In practice, this last step was only needed for AGC~121790, AGC~219199, and AGC~238961. 

    \item We used the task \textsl{specsmooth} to apply Hanning smoothing to suppress any Gibbs ringing, and we decimated the cube by a factor of two, retaining every other channel map to reduce channel-to-channel correlation. This procedure yields a final spectral resolution (and channel width) of $13.06\, \mathrm{kHz}$ ($\Delta_{\mathrm V}\,{\sim}\,2.76\,\mathrm{km\,s^{-1}}$). Finally, we applied a frequency-dependent primary-beam correction using \textsl{impbcor}, after regridding the primary-beam model to the decimated spectral grid, yielding the final science-ready data cube (produced with \textsl{exportfits}). The cube properties of the MeerKAT targets are listed in Table~\ref{tab:beams}.
\end{enumerate}

\begin{table}
\caption{Properties of \textit{FRIDA}'s data cubes.}
\phantomsection
\label{tab:beams}
\centering
\resizebox{\columnwidth}{!}{
\begin{tabular}{ccccc}
	\hline \noalign{\vskip 1.1pt} 
 Name & $b_{\rm size}$ & $b_{\rm PA}$ & $\sigma_{\rm chan}$ & $\Delta_{\rm V}$ \\ \noalign{\vskip 2.4pt}
      & [arcsec$\times$arcsec] & [deg] & [mJy/beam] & [$\mathrm{km\,s^{-1}}$]   \\ \noalign{\vskip 2.4pt}
      (1) & (2) & (3) & (4) &  (5) \\ \noalign{\vskip 2.4pt}
   \hline \noalign{\vskip 1.5pt}  
\noalign{\vskip 1.3pt}
AGC~103435   & $11.05\times8.61$ & 2.6 &  0.16  & 2.76\\ \noalign{\vskip 2.4pt}
AGC~114905   & $9.00\times9.00$ & $0.0$ &  0.26 & 3.40\\ \noalign{\vskip 2.4pt}
AGC~121790   & $11.98\times8.53$ & $-21.4$ & 0.14 & 2.76 \\ \noalign{\vskip 2.4pt}
AGC~189096   & $11.70\times8.97$ & 6.4 & 0.16 & 2.76\\ \noalign{\vskip 2.4pt}
AGC~198477   & $11.78\times8.90$ & 6.5 & 0.15 & 2.76  \\ \noalign{\vskip 2.4pt}
AGC~198540   & $13.18\times9.57$ & $-5.7$   &  0.13  & 2.76   \\ \noalign{\vskip 2.4pt}
AGC~219199  & $13.18\times9.33$ & 22.4 & 0.14 & 2.76\\ \noalign{\vskip 2.4pt}
AGC~226721   & $11.61\times8.95$ & 10.5 & 0.15 & 2.76  \\ \noalign{\vskip 2.4pt}
AGC~238961   & $12.04\times9.10$ & $-5.6$ & 0.15 & 2.76 \\ \noalign{\vskip 2.4pt}
AGC~238984  & $12.51\times9.16$ & 19.2 & 0.15  & 2.76 \\ \noalign{\vskip 2.4pt}
AGC~242019  & $12.00\times12.00$ & $0.0$ & 0.27  & 7.00  \\ \noalign{\vskip 2.4pt}
AGC~258576 & $11.58\times9.09$ & 9.0 & 0.14  & 2.76  \\ \noalign{\vskip 2.4pt}
   \hline
    \end{tabular} }
    \tablefoot{
    (1): Galaxy ID from the Arecibo General Catalogue. 
    (2): Beam size, $b_{\rm maj}\,{\times}\,b_{\rm min}$.
    (3): Beam position angle.
    (4): Average channel rms noise in mJy/beam.
    (5): Spectral resolution (radiovelocity convention); this matches the cube channel width, except for AGC~114905, for which the channel width is $\Delta_{\rm V}/2$. 
    }
\end{table}

\section{\hi properties}
\label{sec:HIprop}

\subsection{Fluxes, \hi masses, and intensity maps}

\begin{table*}
\caption{\hi properties of our galaxy sample.}
\label{tab:properties}
\centering
\begin{tabular}{ccccccccc}
	\hline \noalign{\vskip 1.1pt} 
 Name & $S^{V_{\rm obs}}$ & $D_{\rm H}$ & $M_{\rm HI}$ & $PA$ & $i_{\rm HI}$ & $R_{\rm HI}$ & $R_{\rm 50,HI}$ & $R_{\rm b,HI}$ \\ \noalign{\vskip 2.4pt}
      & $[\mathrm{Jy\, km\,s^{-1}}]$ & $[\mathrm{Mpc}]$ & $[10^{8}\,M_\odot]$ & [deg] & [deg] & [kpc] & [kpc] & [kpc]  \\ \noalign{\vskip 2.4pt}
      (1) & (2) & (3) & (4) &  (5) & (6) & (7) & (8) & (9) \\ \noalign{\vskip 2.4pt}
   \hline \noalign{\vskip 1.5pt}  
\noalign{\vskip 1.3pt}
AGC  103435 & $1.07\pm0.09$  & $28.5\pm3.5$ & $2.0^{+0.5}_{-0.5}$ & $150\pm14$ & $30\pm11$ & $4.5\pm0.6$ &  $2.9\pm0.4$ & $3.3_{-0.5}^{+0.2}$\\ \noalign{\vskip 2.4pt}
AGC  114905 & $0.74\pm0.07$ & $75.7\pm5.2$ & $10.0^{+1.4}_{-1.3}$ & $85\pm12$ & $31\pm4$ & $11.4^{+0.8}_{-0.9}$ & $7.5\pm0.5$ & $8.5_{-1.0}^{+1.6}$ \\ \noalign{\vskip 2.4pt}
AGC  121790 & $2.44\pm0.20$ & $40.0\pm3.9$ & $9.2^{+1.9}_{-1.7}$ & 252$\pm$6 & $37\pm3$ & $9.4\pm0.9$ & $6.1\pm0.6$ & $6.2_{-0.5}^{+0.7}$ \\ \noalign{\vskip 2.4pt}
AGC  189096 & $1.09\pm0.09$  & $60.0\pm4.7$ & $9.2^{+1.5}_{-1.4}$ & $62\pm5$ & $43\pm3$ & $10.3\pm0.9$ & $7.0\pm0.6$ & $7.7_{-0.5}^{+1.0}$  \\ \noalign{\vskip 2.4pt}
AGC  198477 & $2.50\pm0.20$ & $77.9\pm5.3$ & $35.8^{+5.0}_{-4.7}$ & $134\pm6$ & $49\pm4$ & $19.0\pm1.4$ & $12.1^{+1.0}_{-0.9}$ & $13.5_{-2.0}^{+2.3}$ \\ \noalign{\vskip 2.4pt}
AGC  198540 & $1.87\pm0.15$ & $63.4\pm4.8$ & $17.6^{+2.8}_{-2.5}$ & $178\pm12$ & $27\pm5$ & $15.0^{+2.4}_{-3.2}$ & $14.2\pm1.1$ & $19.2_{-1.1}^{+1.9}$\\ \noalign{\vskip 2.4pt}
AGC  219199 & $1.56\pm0.13$ & $39.3\pm3.9$ & $5.7^{+1.2}_{-1.1}$ & $120\pm15$ & $24\pm7$ & $7.3\pm0.7$ & $4.7\pm0.5$ & $5.0_{-0.6}^{+0.3}$  \\ \noalign{\vskip 2.4pt}
AGC  226721 & $1.15\pm0.09$ & $43.5\pm4.0$ & $5.1^{+1.0}_{-0.9}$ & $253\pm8$ & $50\pm5$ & $7.3\pm0.7$ & $4.9\pm0.5$ & $5.3_{-0.7}^{+0.4}$ \\ \noalign{\vskip 2.4pt}
AGC  238961 & $0.85\pm0.07$ & $67.8\pm4.9$ & $9.1^{+1.4}_{-1.3}$ & $61\pm14$ & $43\pm6$ & $9.4^{+0.8}_{-0.7}$ & $6.1^{+0.6}_{-0.5}$ & $5.0_{-0.7}^{+1.2}$  \\ \noalign{\vskip 2.4pt}
AGC  238984 & $0.41\pm0.03$ &$107.0\pm6.3$ & $11.1^{+1.4}_{-1.3}$ & $210\pm10$ & $45\pm7$ & $11.4\pm0.8$ & $8.6^{+1.0}_{-0.8}$ & $7.0_{-1.0}^{+1.9}$  \\ \noalign{\vskip 2.4pt}
AGC  242019 & $3.42\pm0.34$ & $28.9\pm7.5$ & $6.7^{+3.9}_{-3.0}$ & $6\pm4$ & $64\pm3$ & $8.8\pm2.3$ & $6.3\pm1.6$ & $5.9\pm1.0$ \\ \noalign{\vskip 2.4pt}
AGC  258576 & $1.49\pm0.12$ & $79.4\pm5.4$ & $22.1^{+3.1}_{-2.9}$ & $218\pm4$ & $55\pm2$ & $16.5\pm1.2$ & $11.7^{+0.9}_{-0.8}$ & $12.5_{-2.1}^{+1.8}$ \\ \noalign{\vskip 2.4pt}
   \hline
    \end{tabular} 
    \tablefoot{
    (1): UDG name, from the AGC catalogue. 
    (2): \hi flux. 
    (3): Hubble flow distance.
    (4): \hi mass. 
    (5): Position angle. Measured between North and the receding half of the galaxy in the North-to-East direction.
    (6): \hi inclination. 
    (7): \hi radius.
    (8): \hi half-mass radius.
    (9): \hi break or transition radius.
    }
\end{table*}

To measure \hi fluxes and create \hi maps, we first build a mask data cube for each galaxy. The masks should prevent significant noise from being included while preserving faint \hi emission. For this, we use the S+C source finder in the SoFiA-2 software (see \citealt{sofia_1} and \citealt{sofia_2} for details), although we have verified that different masking strategies do not significantly change the parameters derived below.

\looseness=-1
We measure fluxes in the observed LSRK velocity rest frame ($S^{V_{\rm obs}}$) by integrating the \hi global profiles. We work with data cubes in the optical velocity convention. The flux uncertainty includes both statistical and systematic contributions. The former was derived from the rms noise measured in emission-free regions of the primary-beam-corrected channel maps and propagated by summing the per-channel uncertainties in quadrature over the velocity range used to compute the integrated flux (e.g. \citealt{marc_phd}). The systematic uncertainty is associated with the absolute flux calibration and dominates the error in our flux measurements. We assume a flux calibration error of $10\%$ for the VLA data and of $8\%$ for the MeerKAT data \citep{serra2023}. This calibration uncertainty was treated as a multiplicative error and added in quadrature to the statistical uncertainty. Fluxes and their uncertainties are reported in Table~\ref{tab:properties}. In Appendix~\ref{app:fluxes}, we show the good agreement between our fluxes and those recovered from single-dish data.

Next, we compute \hi masses assuming optically-thin discs\footnote{\hi self-absorption can hide up to $20{-}30\%$ of the intrinsic flux in massive edge-on systems \citep{braun2012,peters2017}. Nevertheless, self-absorption appears to be negligible in gas-rich low-mass galaxies \citep{haynes1984}, so it is unlikely to affect our sample.} with the expression \citep{meyer2017}
\begin{equation}
    \dfrac{M_{\rm HI}}{M_\odot} = \dfrac{2.35\times10^5}{(1+z)^2} \, \left(\dfrac{D_{\rm L}}{\mathrm{Mpc}} \right)^2 \, \left( \dfrac{S^{V_{\rm obs}}}{\mathrm{Jy\,km\,s^{-1}}} \right)~,
\end{equation}
with $D_{\rm L}$ the luminosity distance and $z$ the redshift. For consistency with the literature, we adopt ALFALFA flow-model distances $D_{\rm H}$ \citep{alfalfa}, which agree with the comoving distance $D_{\rm C}$ to within $1\%$ at our $z$. Since $D_{\rm L}\,{=}\,(1+z)\,D_{\rm C}\ {\approx}\ (1+z)\,D_{\rm H}$, the final expression to compute the \hi mass becomes
\begin{equation}
    \dfrac{M_{\rm HI}}{M_\odot} = 2.35\times10^5 \, \left(\dfrac{D_{\rm H}}{\mathrm{Mpc}} \right)^2 \, \left( \dfrac{S^{V_{\rm obs}}}{\mathrm{Jy\,km\,s^{-1}}} \right)~.
\end{equation}

Table~\ref{tab:properties} lists $M_{\rm HI}$ for our galaxies. Uncertainties were obtained from Monte Carlo realisations that accounted for errors in the flux and distance. For the distance, also listed in Table~\ref{tab:properties}, we adopt the average of the uncertainty reported by ALFALFA \citep{alfalfa} and the more conservative flow-distance uncertainty scheme proposed by \cite{haubner2025}. While improved stellar masses for \textit{FRIDA} will be derived from our GTC imaging, initial estimates have been derived by \cite{leisman2017} and \cite{janowiecki2019} based on SDSS imaging (Figure~\ref{fig:reffmstar}), yielding approximate gas fractions in the range $f_{\rm gas}=M_{\rm gas}/(M_{\ast}+M_{\rm gas})\sim0.6{-}0.95$, with $M_{\rm gas}=1.36\, M_{\rm HI}$ after correcting for helium.\smallskip

\noindent
Next, we derive total-intensity \hi maps with the software \bba \citep{barolo}, and use them to build column-density maps following
\begin{equation}
\label{eq:columndensity}
\dfrac{N_{\rm HI}(x,y)}{\rm cm^{-2}} =
1.10 \times 10^{24}\,
(1+z)^2\,
\mathrm{mom0}(x,y)\,
\left(
\frac{b_{\mathrm{maj}}\,b_{\mathrm{min}}}
{\rm arcsec^{2}}
\right)^{-1}
\end{equation}
where $\mathrm{mom0}(x,y)$ represents the moment-0 total-intensity map with units $\rm Jy\,beam^{-1}\, km\,s^{-1}$, obtained from the masked data cube. We estimate a representative noise level $\sigma_{N_{\rm HI}}$ using
\begin{equation}
\begin{split}
\frac{\sigma_{N_{\rm HI}}}{\rm cm^{-2}} &=
1.10 \times 10^{21}\, (1+z)^2\,\sqrt{N_{\rm chan,out}} \\
&\times\,
\left(\frac{\sigma_{\rm chan}}{\rm mJy\,beam^{-1}}\right)\,
\left(\frac{\Delta_{\mathrm{V}}}{\rm km\,s^{-1}}\right)\, 
\left(\frac{b_{\rm maj}\,b_{\rm min}}{\rm arcsec^{2}}\right)^{-1}~,
\end{split}
\end{equation}
with $\sigma_{\rm chan}$ and $\Delta_{\rm V}$ as reported in Table~\ref{tab:beams}, and $N_{\rm chan, out}$ the number of independent channels that contribute to the pixels at the edges of the moment-0 maps (which varies due to our masks).

\begin{figure*}
    \centering
    \includegraphics[width=1\linewidth]{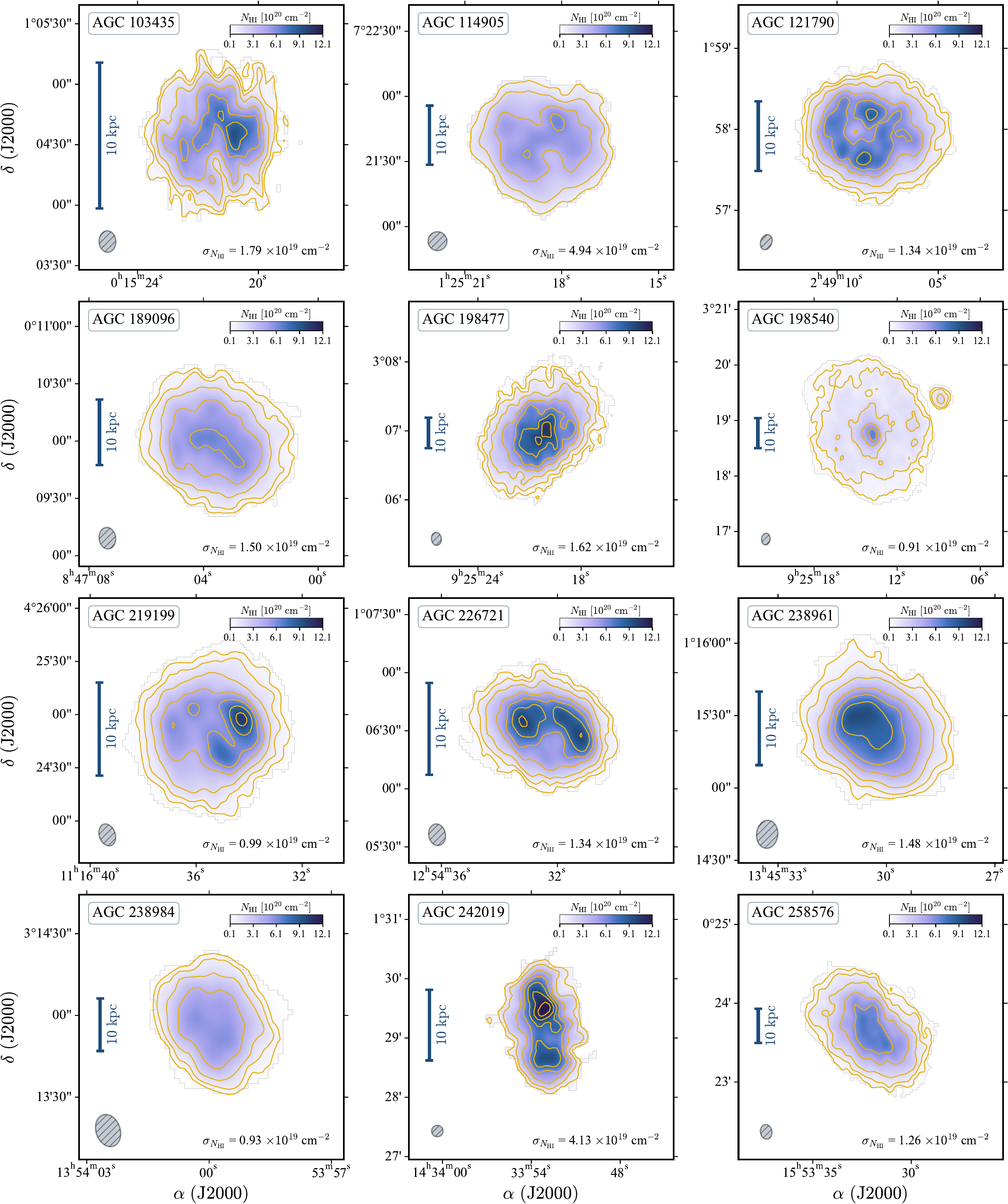}
    \caption{\textit{FRIDA} \hi column-density maps. On each panel, the emission goes down to $1\,\sigma_{N_{\rm HI}}$, with $\sigma_{N_{\rm HI}}$ indicated in the bottom-right corner. The contours are plotted at $N_{\rm HI}=0.5, 1, 2n\,{\times}\, 10^{20}\,\text{cm}^{-2}$, with $n \in \mathbb{Z}_{\geq 1}$, showing only levels at or above $2\,\sigma_{N_{\rm HI}}$. The beam of each galaxy is shown in the bottom-left corner, while the vertical band represents a $10{-}\rm{kpc}$ scale at the distance of each UDG. All galaxies share the same colour-bar scheme, shown in the top right of each panel.
    }
    \label{fig:nhimaps}
\end{figure*}

\looseness=-1
Figure~\ref{fig:nhimaps} shows the column-density maps of our UDGs, down to a $1.5\,\sigma_{N_{\rm HI}}$ level. The \hi emission is extended and fairly symmetric, resembling ordered discs seen at various inclinations. Some patchy substructures can also be seen at high $N_{\rm HI}$, and in upcoming studies we will contrast them against the optical emission of the galaxies. Overall, the similarities and differences in the gas distribution among UDGs will become more evident in Section~\ref{sec:surfden}, where we present their \hi mass surface density radial profiles. It is worth highlighting the case of AGC~198540, which has a markedly different \hi morphology and a satellite companion, both of which we discuss in Section~\ref{sec:a198540}.

\subsection{Inclinations and position angles of the \hi discs}
\label{sec:painc}

We determine the inclination of the \textit{FRIDA} \hi discs using the software \texttt{cannubi} (see \citealt{huds2020,filippo_highz,fernanda2023}), which employs forward modelling to find the position angle ($PA_{\rm HI}$) and inclination ($i_{\rm HI}$) of a model galaxy whose \hi map better matches the observations after taking into account the shape of the beam. To further reduce any effect of the ellipticity of the beam and increase the S/N in the outskirts of the galaxies, we run \texttt{cannubi} on cubes smoothed down to a Gaussian beam with a FWHM of $15\,\mathrm{arcsec}\,{\times}\,15\,\mathrm{arcsec}$, which we find to be a reasonable scale to increase the outer S/N while keeping the galaxies resolved. We assume a disc scale height of 500 pc, typical for gas-rich dwarfs \citep{banerjee2011,ceci_dwarfs,paper_massmodels,paper_galaxyhalo}, but using other plausible values (${\sim}\,0.2{-}1\,\rm{kpc}$, e.g. \citealt{ceci_q3d,agc114905_deep}) does not alter our results.

We adopt the $i_{\rm HI}$ from \texttt{cannubi} as our fiducial inclination. For the position angle $PA$, we follow a hybrid approach. We first input the best-fitting \texttt{cannubi} $PA_{\rm HI}$ as the initial guess for our kinematic modelling (Mancera Piña et al., in prep), and use \bba to find the kinematic position angle $PA_{\rm kin}$. We adopt as $PA$ the resulting $PA_{\rm kin}$, and its uncertainty is the largest value among the errors found by \texttt{cannubi} and \bba. For all but one UDG, the $PA_{\rm HI}$ agrees with $PA_{\rm kin}$ within the uncertainties. The exception is AGC~103435, whose morphological and kinematic position angles are offset by more than $60^\circ$. The optical image and velocity field of the galaxy exhibit a disturbed morphology, which we will further discuss in our upcoming studies and may be a consequence of recent gas accretion (e.g. \citealt{casanueva2026}) or stellar activity. For this UDG, we adopt $PA_{\rm HI}$ (which aligns well with the optical morphology) as our fiducial $PA$. The inclinations and position angles of our sample are listed in Table~\ref{tab:properties}. Two galaxies, AGC~198477 and AGC~198540, have a warp in $PA_{\rm kin}$ of about $15^\circ$; for them, the entry in Table~\ref{tab:properties} corresponds to the average value of $PA_{\rm kin}$ (consistent with $PA_{\rm HI}$ within $1\,\sigma$), and the uncertainty is the largest value between the standard deviation in $PA_{\rm kin}$ and the error in $PA_{\rm HI}$.
In Appendix~\ref{app:bias_inclination}, we show the good agreement between the \hi total-intensity maps at 15-arcsec resolution and the \texttt{cannubi} models with our adopted position angles and inclinations. 

As reported in Table~\ref{tab:properties}, the inclinations of \textit{FRIDA}'s gas-rich UDGs have a median of $43^\circ$. This differs from the expected median of $60^\circ$ for a sample of disc galaxies randomly oriented. In Section~\ref{sec:bias_inclination_discussion}, we examine this in detail, showing that the inclinations reported in this work are fully consistent with expectations once selection effects are taken into account.

\subsection{Gas surface density profiles}
\label{sec:surfden}
We derive gas surface density profiles $\Sigma_{\rm HI}(R)$ as follows. First, we use our column-density maps (Eq.~\ref{eq:columndensity}) and convert them to face-on surface density maps through
\begin{equation}
    \dfrac{\Sigma_{\rm HI}(x,y)}{M_\odot\,\mathrm{pc}^{-2}} = \dfrac{N_{\rm HI}(x,y)}{1.25\times10^{20}\,\mathrm{cm^{-2}}} \,\cos(i_{\rm HI})~.
\end{equation}

\noindent
Next, we extract azimuthal radial profiles $\Sigma_{\rm HI}(R)$ by averaging the surface density maps using concentric elliptical rings oriented according to our $PA$ (tracing the warp if present) and $i_{\rm HI}$. We sample the profiles using rings separated by $\sqrt{b_{\rm maj}b_{\rm min}}/2$ arcsec, corresponding to approximately Nyquist sampling of the beam-convolved radial profile while limiting the correlation between adjacent measurements. Uncertainties were propagated using a moment-0 error map derived from masked data cubes, accounting for the channel rms noise and spectral resolution reported in Table~\ref{tab:beams}. To incorporate uncertainties in the disc geometry, we performed a Monte Carlo propagation in which global offsets $\Delta {PA}$ and $\Delta i_{\rm HI}$ were drawn from Gaussian distributions with standard deviations corresponding to the uncertainties reported in Table~\ref{tab:properties}. These offsets were applied uniformly at all radii, and for each realisation, the surface density profile was re-extracted using the perturbed geometry. In addition, random noise consistent with the annular mean uncertainty derived from the moment-0 error map was added to each radial bin. The adopted profile corresponds to the median of the resulting ensemble of realisations (which closely matches the default original realisation), with asymmetric uncertainties defined by the 16th and 84th percentiles. Finally, an absolute flux-scale uncertainty was propagated as a correlated multiplicative term by applying a global random scaling factor to each realisation ($10\%$ for VLA data and $8\%$ for MeerKAT data). Our $\Sigma_{\rm HI}$ profiles can be downloaded from \textit{FRIDA}'s website. 

From the $\Sigma_{\rm HI}$ profiles we also extract (using linear interpolation if needed) two additional informative quantities: $R_{\rm HI}$, the radius at which $\Sigma_{\rm HI}(R)\, {=}\, 1\, M_\odot\,\mathrm{pc^{-2}}$ (often called the \hi radius), and $R_{\rm 50, HI}$, the radius enclosing $50\%$ of $M_{\rm HI}$. These quantities will be further discussed in the following section and are listed in Table~\ref{tab:properties}, together with their uncertainties. As before, the quoted errors are derived from Monte Carlo realisations and include contributions from flux, geometry, and distance uncertainties. Some studies (e.g. \citealt{wang16,lexi}) apply an approximate beam correction to the observed \hi sizes, e.g. $R_{\rm HI}^2\, {=}\, R_{\rm HI,obs}^2\, {-}\, (b_{\rm maj}\,b_{\rm min}/4)$. We have not applied this correction, which is negligible (${\sim}\,1\%$) at our high spatial resolution.

\begin{figure*}
    \centering
\includegraphics[width=1\linewidth]{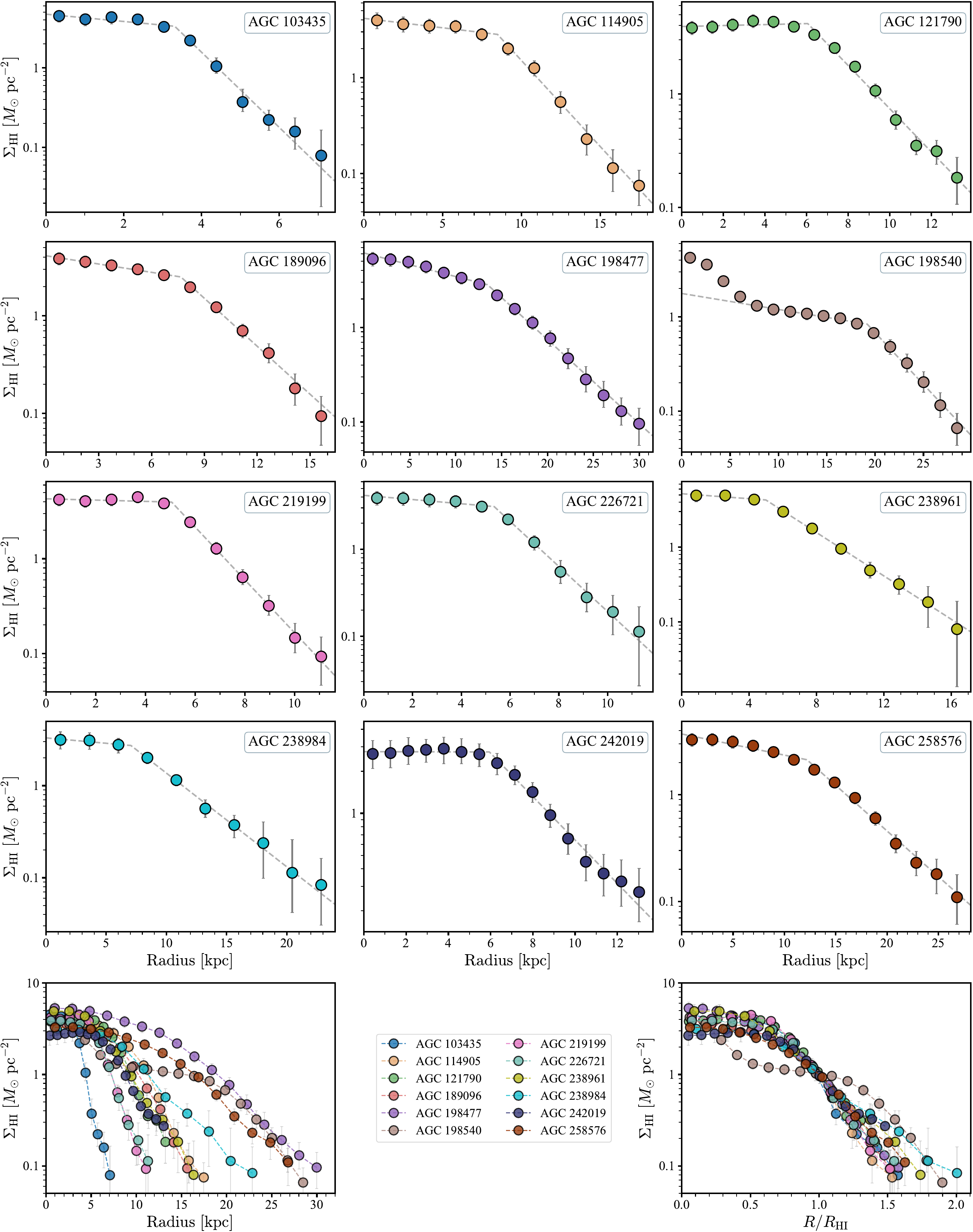}
    \caption{\hi surface density profiles for the \textit{FRIDA} UDGs. The grey dashed lines in the single-galaxy panels represent a broken exponential profile fitted to the data.
    The bottom panels show all the profiles together (dashed colour lines connecting measurements from the same galaxy) as a function of $R$ and $R/R_{\rm HI}$.
    }
    \label{fig:sigma_profs}
\end{figure*}

Figure~\ref{fig:sigma_profs} shows the \hi density profiles of our galaxies. The upper four rows present each galaxy in individual panels, while the bottom row shows all the galaxies together as a function of $R$ in kiloparsec units and in units of $R_{\rm HI}$. The azimuthal profiles are fairly smooth and highlight the large extent of the gas discs, with some having diameters of $20{-}30$ kpc.
The profiles have a fair degree of similarity, showing an inner plateau with a central maximum value of $\Sigma_{\rm HI}(R{\rightarrow}0)\,{\approx}\,4\,{\pm}\,1\, M_\odot\,\rm{pc^{-2}}$, followed by an exponential decline in the outer parts. This type of profile is characteristic of gas-rich galaxies \citep{swaters02,wang16}, and we will delve into further comparisons in Section~\ref{sec:selfsimilar}. We find that a broken exponential function fits the data very well, as shown in Figure~\ref{fig:sigma_profs} with grey dashed lines. The transition or break radius of the broken exponential ($R_{\rm b,HI}$) is listed in Table~\ref{tab:properties}, and has a median value of $0.72\,R_{\rm HI}$ or $1.05\,R_{\rm HI,50}$; we plan to return to this quantity in future work, when also studying the edges of the stellar discs of our sample. Beyond $R_{\rm b,HI}$, the profiles decline exponentially, with a median scale length of $0.26\,R_{\rm HI}$. The only galaxy with a clearly different profile is AGC~198540 (further discussed in Section~\ref{sec:a198540}), which has a central density at a similar level to the other UDGs but has a steep central profile within the inner 8 kpc that then flattens and declines further out, as also seen in its column-density map in Figure~\ref{fig:nhimaps}.

\section{Results and discussion}
\label{sec:mainresults}

\subsection{The inclination angles of gas-rich UDGs}
\label{sec:bias_inclination_discussion}

The distribution of inclination angles in \textit{FRIDA} UDGs has a median value of $43^\circ$, in resemblance to those reported from low-resolution data in other gas-rich UDGs by \cite{huds2019} and \cite{lexi} and in contrast with the median value of $60^\circ$ expected for a population of galaxies viewed at random angles (see e.g. \citealt{romanowsky} and below). 

This is not surprising, as selection effects are expected to change this distribution (e.g. \citealt{jones1996}). 
First, \hi surveys such as ALFALFA can be biased against edge-on galaxies because, at a fixed \hi flux and rotational speed, their emission spreads across more spectral channels, lowering their per-channel S/N and thus their detectability.
Second, since UDGs are selected using surface-brightness cuts, one expects a bias toward lower inclinations. This is because a galaxy of a given intrinsic luminosity, viewed at low inclinations, will appear fainter than at higher inclinations due to emission overlap along the same line of sight in the latter case, making it more likely to be classified as a UDG. On the other hand, however, stellar mass functions show that low-mass, low-surface-brightness galaxies are far more numerous than bright high-mass systems (e.g. \citealt{bothun1997,kelvin2014}), which can bias the median inclination high when surface brightness cuts are applied: Although individual galaxies are dimmed when viewed face-on, the much larger reservoir of faint systems means that a non-negligible fraction may only cross the detection threshold when viewed at higher inclinations, (Eddington-) biasing the observed inclination distribution toward more edge-on systems. Quantifying these competing biases is crucial, as it has repercussions not only for the normalisation of the \hi mass-size relation, but also for the upcoming kinematic and dynamic analyses of \textit{FRIDA}.

\begin{figure}
        \includegraphics[width=\linewidth]{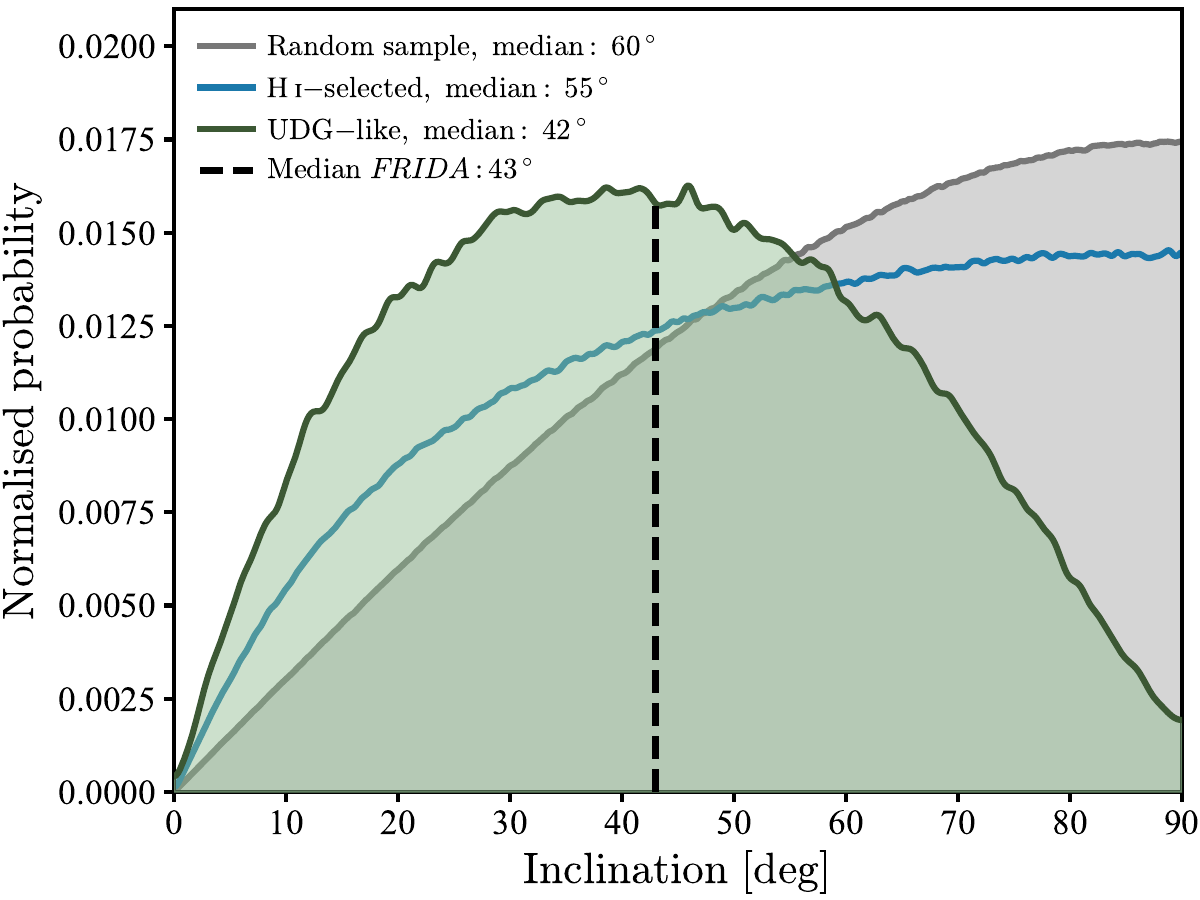} 
        \caption{Normalised probability of inclination angles. The grey distribution represents a randomly inclined disc population, the blue distribution represents an \hi-detected subsample, and the green distribution represents a gas-rich UDG-like population.
        The median inclination in \textit{FRIDA} is $43^\circ$ (black dashed line), consistent with the expectations after accounting for selection effects.
        }
        \label{fig:inclinations}
\end{figure}

To estimate the inclination bias present in our sample, we create a synthetic galaxy population as follows. First, we sample a Schechter stellar mass function with parameters given by \cite{kelvin2014} for Sd-Irr galaxies (a slope of $-1.36$ and a characteristic knee at $10^{9.57}\, M_\odot$).
Next, we assign a stellar effective radius to each galaxy by randomly sampling a Gaussian distribution centred at the $M_\ast{-}R_{\rm e}$ relation from \cite{lange2015} and with a standard deviation in $\log(R_{\rm e})$ of 0.15 dex \citep{fernandezlorenzo,lange2015,ward2024}. Furthermore, we assume exponential profiles (a fair approximation for UDGs, e.g. \citealt{udgs_kiwics,smudges_iii}), which, together with the above parameters, fully define $I(R)$, the face-on surface mass density profile of our mock galaxies. Next, we convert $I(R)$ into an $r$-band $\mu(R)$ assuming a solar absolute magnitude of $4.65\,\mathrm{mag}$ \citep{magnitude_sun} and a mass-to-light ratio $M/L_{r}\,{=}\,0.77$, following the calibrations from \cite{herrmann_m2l} and \cite{du2020} for a colour $g{-}r\,{=}\,0.35$, typical of field LSB galaxies (e.g. \citealt{fernandezlorenzo2012,leisman2017}). Finally, we project our face-on $\mu(R)$ into an on-sky observed $\mu(R)_{\rm obs}$. To incorporate discs of realistic thickness ($q_0\,{\approx}\,0.25$) rather than razor-thin discs, we follow \cite{trujillo2020} and consider the inclination correction $\mu(R)_{\rm obs}\, {=}\, \mu(R)\, {-}\, \Delta_{\mu}$, with $\Delta_\mu\, {=}\, 2.845\, {-}\, 7.833\cos(i) + 10.792\cos(i)^2 - 8.482\cos(i)^3 + 2.679\cos(i)^4$, although we note that using $q_0\,{=}\,0$ or $q_0\,{=}\,0.4$ introduces only a minor shift of ${\sim}\,1^\circ$ to our results below. For $i$, we randomly sample a uniform distribution over $\cos(i)$, as expected for random inclinations in an axisymmetric geometry. The underlying assumption here is that observed galaxies are circular discs projected at a given inclination angle (i.e. the orbits are circular and not elliptical, cf. \citealt{marasco_barsinDM,banik2022}). The resulting disc population, with a median of $60^\circ$, is shown as a grey distribution in Figure~\ref{fig:inclinations}, which shows the normalised probability of selecting galaxies of a given inclination.

Next, we simulate the inclination-dependent selection effects of the ALFALFA survey. The observed ALFALFA signal-to-noise ratio scales approximately as S/N$\,{\propto}\,(\sin i)^{-1/2}$ \citep{alfalfa40}, favouring the detection of more face-on galaxies at fixed integrated \hi flux. We assign each mock galaxy an edge-on S/N drawn from a log-normal distribution calibrated through forward modelling to reproduce the observed ALFALFA S/N distribution. This calibration accounts for isotropic inclinations, the inclination dependence of S/N, and the gradual completeness transition. We apply the same prescription to our mock population and retain objects with measured S/N$\,{\geq}\,6.5$, the ALFALFA threshold for high-quality detections, obtaining an \hi-selected population. Plausible variations in the S/N calibration change the median selected inclination by only ${\sim}\,1{-}2^\circ$.

The inclination probability distribution of this \hi-selected population is shown with a blue curve in Figure~\ref{fig:inclinations}. While the fraction of edge-on galaxies is lower than in the random case, the median is still $54^\circ$, which also matches the median optical inclination found by \cite{alfalfa_sdss} in the ALFALFA-SDSS catalogue ($55^\circ\,{-}\,60^\circ$ for reasonable values of $q_0$). The impact of the \hi selection is systematic but of relatively modest magnitude. This is qualitatively consistent with detectability tests based on ALFALFA spectra of \textit{FRIDA} galaxies, as described in Section 6.2 in \cite{barbara_udgs}, which shows that 11 out of the 12 UDGs in \textit{FRIDA} would have been detected even if edge-on.

Finally, we perform optical selection cuts. \textit{FRIDA} UDGs are drawn from the ALFALFA parent samples by \cite{leisman2017} and \cite{janowiecki2019}, who used the selection criterion $R_{\rm e}\, {\geq}\, 1.5\,\mathrm{kpc}$ and $\langle \mu(R_{\rm e})\rangle_{r}\, {\geq}\, 24\,\mathrm{mag\,arcsec{^{-2}}}$. We therefore apply the same selection cuts to our synthetic population. Furthermore, we discard galaxies with $\langle \mu(R_{\rm e})\rangle_{r}\, {\geq}\, 27.5\,\mathrm{mag\,arcsec{^{-2}}}$, to account for typical detection limits in optical observations of UDGs (e.g. \citealt{leisman2017,udgs_kiwics,smudges_iii}).

The inclination distribution of the selected mock UDG population is displayed in green in Figure~\ref{fig:inclinations}. The UDG-like distribution peaks at intermediate inclinations, with a median of $42^\circ$, and is fairly flat from ${\sim}\,35^\circ{-}50^\circ$ before dropping sharply toward higher and lower inclinations. For reference, the distribution of inclination angles after applying the surface-brightness cut alone has a similar shape and a median of $48^\circ$, which we would expect for disc UDGs not \hi-selected.

Two important implications derive from our analysis. First, the inclinations measured in this work and in \cite{huds2020}, based on \hi morphologies, are remarkably consistent with being drawn from the mock gas-rich UDG distribution. This illustrates that we preferentially sample inclinations around $45^\circ$ due to the selection criteria for UDGs. Second, while there is a bias favouring lower-than-random inclinations, its effect is not strong enough to lead us to select only face-on galaxies. For example, \cite{lelli2024} suggested that the six UDGs outliers of the BTFR from \cite{huds2019} could be reconciled with the BTFR if their inclinations were overestimated. As discussed in \cite{huds2019}, returning their UDGs to the BTFR would require all their inclinations to lie between $10^\circ{-}20^\circ$, which is highly unlikely given the findings in Figure~\ref{fig:inclinations}. For instance, the probability that all six UDGs have $i\,{<}\,15^\circ$ is ${\sim}\,5\,{\times}\,10^{-7}$. The \hi inclinations in \textit{FRIDA} will be double-checked with the optical inclinations derived from our ultra-deep GTC imaging (e.g. \citealt{agc114905_deep}) in an upcoming study; preliminary results in about half the sample show excellent agreement.

\subsection{The \hi mass-size relations of \textit{FRIDA} UDGs}

 \begin{figure*}
\centering
   \includegraphics[width=17cm]{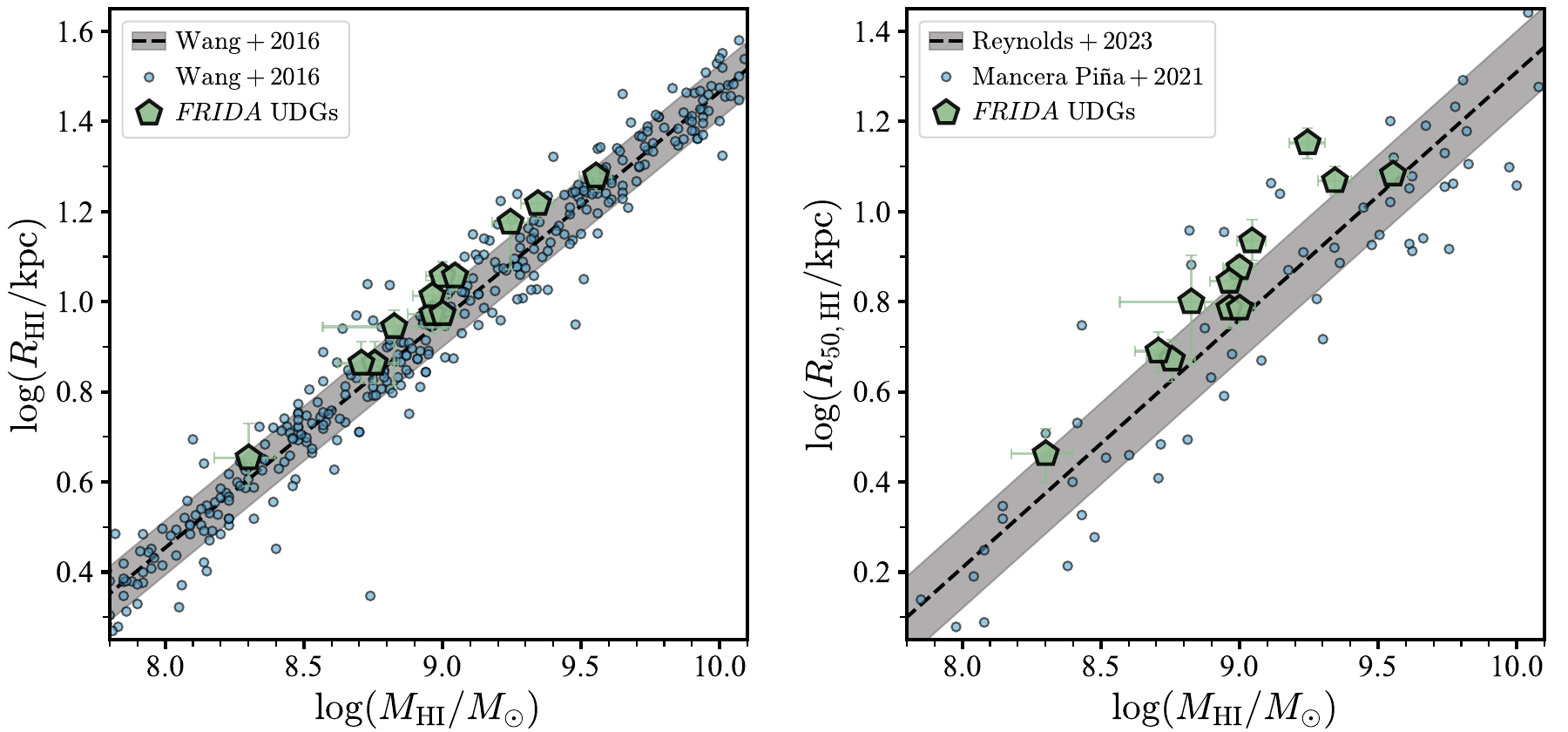}
     \caption{$M_{\rm HI}{-}R_{\rm HI}$ (left) and $M_{\rm HI}{-}R_{\rm 50,HI}$ (right) relations for the UDGs in \textit{FRIDA}. The left panel compares \textit{FRIDA} with measurements and the best-fit relation from \cite{wang16}, while the right panel compares against measurements from \cite{paperIBFR} and the best fit from \cite{reynolds2023}. Grey-shaded bands show the intrinsic scatter of the relations. In both panels, we have applied a small ($-0.04$ dex) horizontal offset to AGC~238961 to avoid overlapping with AGC~121790. Both panels have the same vertical and horizontal spans.}
     \label{fig:mass_size}
\end{figure*}

The \hi mass-size relation ($M_{\rm HI}{-}R_{\rm HI}$) connects the total \hi mass of a galaxy with its isophotal radius enclosing $1\, M_\odot\,\mathrm{pc^{-2}}$ \citep{broeils1997}. Observations and simulations show that this relation holds for all gas-rich galaxies with remarkably low scatter \citep{wang16,rajohnson2022}, unless the discs are heavily disrupted due to strong environmental effects or under overly strong feedback prescriptions \citep{e_wang2015,stevens_masssize,hank2025,rabyang2026}. The tightness and universality of the relation are thought to result from dwarf and massive disc galaxies having fairly homogeneous exponential outer profiles \citep{swaters02,bigiel2012,wang2014,wang16,stevens_masssize}. The similarity in the outer parts of galaxies is perhaps not surprising, as it is set by the balance among dust shielding, the UV background, and metallicity, which are arguably similar among disc galaxies. 

Studying the location of gas-rich UDGs in the $M_{\rm HI}{-}R_{\rm HI}$ plane is interesting as some of their proposed formation mechanisms suggest specific feedback prescriptions leading to bursty star formation histories \citep{nihao_udgs}, more efficient gas accretion \citep{agc114905_deep}, or early major mergers \citep{wright_udgs}, all of which could affect the gas reservoirs and disc structure and therefore the $M_{\rm HI}{-}R_{\rm HI}$ relation. 

The only study of UDGs in the $M_{\rm HI}{-}R_{\rm HI}$ plane has been carried out by \cite{lexi}, who analysed VLA observations of 11 systems and found that they follow the same relation as other disc galaxies (see also \citealt{zhou2026}). However, the $R_{\rm HI}$ measurements in that study were subject to significant uncertainties, as the VLA data were only barely spatially resolved (${\sim}\,1{-}1.5$ independent resolution elements per galaxy side) and could be affected by beam effects. With the high-resolution data of \textit{FRIDA}, we are now able to obtain for the very first time detailed \hi surface density profiles for a sample of gas-rich UDGs, enabling an accurate determination of their position with respect to the established $M_{\rm HI}{-}R_{\rm HI}$ relation. In addition to $R_{\rm HI}$, we examine $R_{\rm 50, HI}$, which is more sensitive to the inner mass distribution.

Figure~\ref{fig:mass_size} presents the \hi mass-size planes. The left panel shows the $M_{\rm HI}{-}R_{\rm HI}$ plane, and we compare our UDGs against the measurements and best-fitting power-law ($\log(R_{\rm HI}/\mathrm{kpc})\,{=}\,0.506\log(M_{\mathrm{HI}}/M_\odot)\, {-}\, 3.594$) from \cite{wang16}. The data from \cite{wang16} have a wider $M_{\rm HI}$ span, but here we focus on a narrower range to appreciate the location of our UDGs better. The UDGs in \textit{FRIDA} lie close to the established relation, consistent with the early results in \cite{lexi}. However, our new data reveal that UDGs systematically lie above the best-fitting relation, with a mean offset in $\Delta\log(R_{\rm HI}/\rm{kpc})\,{=}\,0.06$ dex (${\approx}\,15\%$). This offset is moderate and matches the reported $1\,\sigma$ scatter of the relation (${\sim}\,0.04{-}0.06\,\rm{dex}$, \citealt{wang16,reynolds2023}), but is systematic. To quantify its significance, we use a Monte Carlo test, randomly drawing 12 galaxies from the sample by \cite{wang16} within our $M_{\rm HI}$ range. Out of $10^6$ realisations, only $0.1\%$ produce a mean offset at least as large as the observed value. Moreover, only in $0.03\%$ of the realisations do the randomly selected galaxies lie above the relation. 
Our Monte Carlo tests show thus that the observed offsets are unlikely to arise from random sampling.

A similar behaviour is seen in the right panel of Figure~\ref{fig:mass_size}, where we place the UDGs in the $M_{\rm HI}{-}R_{\rm 50, HI}$ plane, and compare them against a measurement set derived from the data by \cite{paperIBFR} and the best-fitting relation ($\log(R_{\rm HI,50}/\mathrm{kpc})\,{=}\,0.55\log(M_{\mathrm{HI}}/M_\odot)\, {-}\, 4.19$) by \cite{reynolds2023}. Again, all of our UDGs lie slightly above the mean relation, with a median offset $\Delta\log(R_{\rm 50, HI}/\rm{kpc})\,{=}\,0.10$ dex (${\approx}\,25\%$). Performing a similar Monte Carlo test as before with the galaxies from \cite{paperIBFR}, we find that a similar offset is achieved in about $0.7\%$ of the realisations, and the randomly selected galaxies lie all above the best-fit relation only $0.002\%$ of the time. 

To ensure that the offset observed in Figure~\ref{fig:mass_size} is real and not a systematic effect of our method, we verify the recovery of the mass-size relations using a control sample. Specifically, we consider gas-rich galaxies with kinematic models from \cite{paper_galaxyhalo} (since their geometric parameters are derived similarly to ours) and follow the same procedure to derive $R_{\rm HI}$ and $R_{\rm 50, HI}$. We find that the control sample (not shown) scatters around the reference relations in Figure~\ref{fig:mass_size} without clear systematic offsets, suggesting that our analysis does not introduce a general bias in the derived \hi sizes.

We also note that our sample angular-size selection criterion (see Section~\ref{sec:data}) was based on diameters predicted from ALFALFA \hi masses, rather than measured diameters, and therefore does not directly select galaxies with positive size residuals.
In fact, \hi-compact UDGs should be easier to detect and overrepresented in galaxy samples. It is also worth to highlight that the ratio between stellar and \hi sizes is roughly constant in disc galaxies (e.g. \citealt{broeils1997,noordermeer}, cf. \citealt{bosma2017} and \citealt{chamba2024}), so one would expect UDGs (which have large $R_{\rm e}$, Figure~\ref{fig:reffmstar}) to have a higher $R_{\rm HI}$ at fixed $M_\ast$, but not necessarily at fixed $M_{\rm HI}$, where they are compared against brighter high-surface-brightness (HSB) galaxies.

Besides having very low scatter, the \hi mass--size relations have a nearly-quadratic behaviour, $M_{\rm HI}\,{\propto}\, R_{\rm HI}^{2}$ and $M_{\rm HI}\,{\propto}\, R_{\rm 50,HI}^{1.8}$ \citep{wang16,reynolds2023}. This implies that galaxies share a similar characteristic \hi surface density, (e.g. $\Sigma_{\rm HI,c}\,{\propto}\, M_{\rm HI}/R_{\rm HI}^{2}$) that varies little from galaxy to galaxy. 
Compared to galaxies sitting on top of the fits from \cite{wang16} and \cite{reynolds2023}, \textit{FRIDA} UDGs have average characteristic surface densities within $R_{\rm HI}$ and $R_{\rm 50, HI}$ lower by about $25\%$ and $35\%$, respectively. Therefore, the results from \textit{FRIDA} suggest that gas-rich UDGs have \hi discs that are slightly more extended and of slightly lower surface density than those of average disc galaxies on the \cite{wang16} and \cite{reynolds2023} relations. Our results are in good agreement with those reported in the pioneering study by \cite{thijs1993}, which found that LSB galaxies tend to have lower average \hi surface densities than HSB systems.

Our findings set a clear test for the \hi properties of UDGs in hydrodynamical simulations. While the location of simulated UDGs in the $M_{\rm HI}{-}R_{\rm HI}$ relation has not been reported in the literature, it is interesting to note that the gas-rich UDGs in the Romulus25 simulations \citep{wright_udgs} appear to show a similar offset as our UDGs, at least qualitatively (A. Wright, priv. comm., 2026). UDGs in Romulus25 have a relatively high spin parameter driven by their latest major mergers, which occur early, typically at $z\,{\geq}\,1$. On the other hand, \cite{nihao_udgs} report that NIHAO UDGs, formed through episodic stellar feedback, have large \hi sizes for their $M_\ast$, but it remains to be seen if this holds at fixed $M_{\rm HI}$. In general, it will be important to contrast our results with mock observations of different simulations that form UDGs through various mechanisms and with different ISM physics prescriptions.

\subsection{Self-similarity of \hi profiles}
\label{sec:selfsimilar}

\begin{figure*}
\centering
\includegraphics[width=1\linewidth]{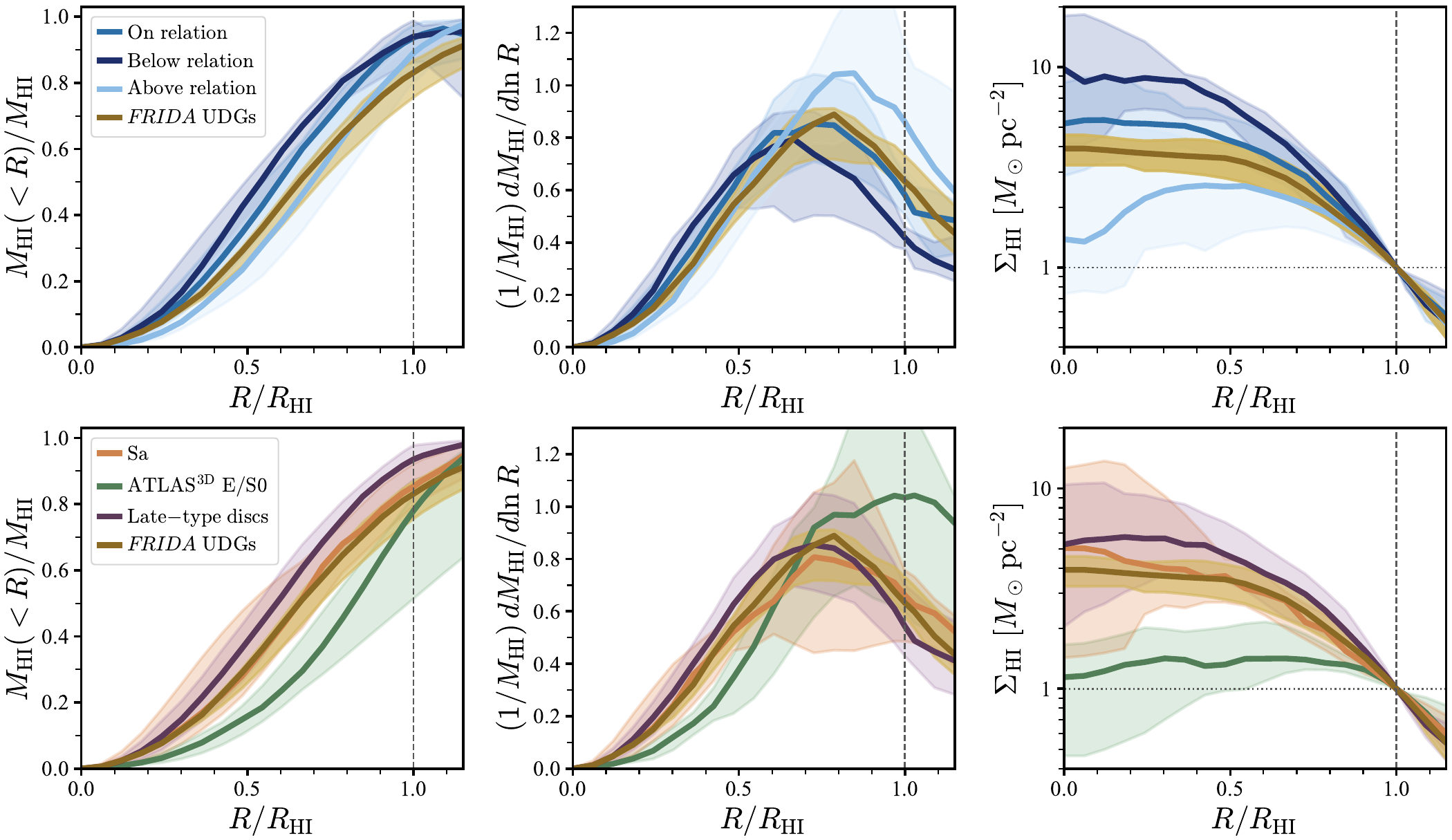}
\caption{\hi distribution within \textit{FRIDA} UDGs compared to galaxies with $8.6\, {\leq}\, \log(M_{\rm HI}/M_\odot)\, {\leq}\, 9.6$ from \cite{wang16}. The left, middle, and right panels show the cumulative \hi mass fraction, the differential \hi mass profile normalised by the total \hi mass, and the \hi surface density, respectively, all as a function of $R/R_{\rm HI}$. The top panels split the comparison sample into galaxies within (blue), above (light blue), or below (dark blue) the $M_{\rm HI}{-}R_{\rm HI}$ relation, using a 0.04 dex threshold. The bottom panels instead split the comparison sample into late-type discs (purple), Sa early-type spirals (orange), and E/S0 early-type galaxies (green). UDGs are shown in dark yellow in both rows. In all panels, the solid curves are the median stacks of individual profiles, and the bands represent the 16th--84th percentile range. Grey reference lines indicate $R=R_{\rm HI}$ and, in the right panels, $\Sigma_{\rm HI}=1\,M_\odot\,{\rm pc}^{-2}$. At $R\,{\geq}\,1.1\,R_{\rm HI}$, the comparison samples become significantly incomplete, so trends beyond that radius should be interpreted with caution.}
\label{fig:cumulative}
\end{figure*}

Given the approximate self-similarity of gas discs reported in the literature, the tendency of UDGs to lie above the mean $M_{\rm HI}{-}R_{\rm HI}$ and $M_{\rm HI}{-}R_{\rm 50,HI}$ relations raises the question of whether their normalised \hi profiles differ from those of the broader population of galaxies. This is also relevant because inspection of the sample from \cite{wang16} reveals that the galaxies above UDGs in the $M_{\rm HI}{-}R_{\rm HI}$ plane are early-type systems, whose $\Sigma_{\rm HI}(R)$ profiles differ from those of late-type discs \citep{swaters02,wang16}.

In Figure~\ref{fig:cumulative}, we compare cumulative \hi mass fractions, differential \hi mass profiles normalised by the total \hi mass, and $\Sigma_{\rm HI}(R)$, all plotted against $R/R_{\rm HI}$. We again use the \hi surface-density profiles and masses from \cite{wang16} and classify the comparison sample using two separate schemes. First, we split the galaxies into those within 0.04 dex of the mean $M_{\rm HI}{-}R_{\rm HI}$ relation and those offset by more than 0.04 dex in either direction. Second, we divide them into late-type discs, Sa spirals, and E/S0 early-type galaxies, using the classifications provided by \cite{wang16}. The E, S0, and Sa galaxies come from the ATLAS$^{\rm 3D}$ \citep{serra2012} and WHISP surveys \citep{whisp,noordermeer}; although this sample is not as comprehensive as the late-type sample, it provides useful comparisons \citep{wang16}. For both schemes, we consider only galaxies with $8.6 \leq \log(M_{\rm HI}/M_\odot) \leq 9.6$, covering the main mass range of our sample.

The top panels of Figure~\ref{fig:cumulative} show progressively less centrally concentrated \hi distributions towards larger positive offsets from the $M_{\rm HI}{-}R_{\rm HI}$ relation. \textit{FRIDA} UDGs lie between galaxies close to the reference relation and those on its upper envelope. 
Differences in the cumulative \hi mass fractions are most pronounced in the inner disc, reaching $20{-}25\%$ between the comparison subsamples at $R\,{\approx}\,0.5\,R_{\rm HI}$, and largely diminish near $R_{\rm HI}$. This explains why the offset is larger in the half-mass-radius relation: the gas is distributed further out within the measured disc, without a correspondingly large excess of detected gas beyond $R_{\rm HI}$. AGC~198540 is an exception owing to its diffuse outer component, discussed in Section~\ref{sec:a198540}.

The morphological comparison in the bottom panels shows that late-type discs have more centrally concentrated \hi distributions and higher surface densities at a given $R/R_{\rm HI}$ than Sa and E/S0 galaxies. \textit{FRIDA} UDGs occupy an intermediate position. Their cumulative \hi mass profiles overlap closely with those of Sa spirals, although their \hi surface densities are lower within the inner ${\sim}\,0.3\,R_{\rm HI}$. Similar results were found by \cite{deblok1996} when comparing LSB and early-type HSB disc galaxies. Beyond ${\sim}\,R_{\rm HI}$, the UDG surface-density profiles resemble those of late-type discs more closely, declining more steeply than the early-type profiles. Hints of an even sharper outer decline or truncation in UDGs remain tentative, because the comparison subsamples become significantly incomplete beyond ${\sim}\,1.1\,R_{\rm HI}$ and the observations have different sensitivities.

\subsection{The curious case of AGC~198540}
\label{sec:a198540}
As mentioned throughout the text, the \hi distribution of AGC~198540 differs significantly from that of the rest of the sample. Specifically, AGC~198540 has a bright and steep centre surrounded by a faint envelope (Figures~\ref{fig:nhimaps} and \ref{fig:sigma_profs}). This envelope is very extended, and inspection of DECaLs optical images (Figure~\ref{fig:stamps}) reveals that the \hi disc of AGC~198540 appears to be overly extended relative to its stellar disc when compared to all the other UDGs in \textsl{FRIDA}, and this will be further analysed with our upcoming GTC imaging.

The above characteristics make AGC~198540 the most extreme of the UDGs in \textit{FRIDA} in terms of its large $R_{\rm 50, HI}$ relative to the mean literature relation. Within $R_{\rm HI}$, the galaxy is also atypical. For the rest of the sample, the average fraction of the \hi flux within $R_{\rm HI}$ is $0.83$ with a scatter of $0.04$. Instead, for AGC~198540 this fraction is $0.55^{+0.11}_{-0.17}$, which is more similar to that of early-type galaxies, which have $0.62\,{\pm}\,0.14$, as reported by \cite{lexi} for ATLAS-3D galaxies \citep{serra2012}. \cite{wang16} has argued that early-type galaxies might have extended \hi distributions and low \hi flux fractions within $R_{\rm HI}$ due to recent gas accretion. 

Remarkably, our deep MeerKAT observations reveal that AGC~198540 has a small and previously unknown companion. The difference in systemic velocities between the UDG and its companion is ${\sim}\,60\,\rm{km\,s^{-1}}$, with a projected distance of about 25 kpc. The companion has $M_{\rm HI}\,{\sim}\,5\,{\times}\,10^7\, M_\odot$ (i.e. around $1/35$ of the \hi mass of AGC~198540) and a blue optical counterpart that is not catalogued in astronomical databases but is visible in DECaLS imaging (see Figure~\ref{fig:stamps}) and will be better characterised by our GTC photometry. Given the much lower \hi mass and smaller size of the companion, we consider it a newly discovered satellite of AGC~198540, although we cannot independently establish if the system is in a bound orbit. Recently, two pairs of gas-rich UDGs have been identified using \hi observations \citep{barbara_udgs,obeirne2025}, and two candidate satellites of gas-poor UDGs have been reported \citep{muller2023}. Yet, to the best of our knowledge, this is the first time that a satellite candidate is found around a gas-rich UDG.

One may wonder whether gas stripped from the companion and subsequently accreted by AGC~198540 could explain its distinct morphology and extended outer envelope, as well as the warp in the kinematic position angle in the outer parts of its disc (Mancera Piña et al., in prep.). Another possible perturber is UGC~5027 (Appendix~\ref{app:neighbours}), located at a projected separation of approximately 330~kpc and differing in systemic velocity by about $85\,{\rm km\,s^{-1}}$. A past fly-by encounter with this galaxy could therefore also have contributed to the morphology of AGC~198540.

The regular kinematics of AGC~198540, with no strong evidence of non-circular motions (Mancera Piña et al., in prep.), place further constraints on the nature of any potential interaction. The encounter with the satellite may not yet have begun, in which case no gas transfer has taken place, and the satellite would not be responsible for the unusual morphology. The interaction (with either galaxy) may also have been recent but relatively mild and therefore left little imprint on the kinematics of AGC~198540. Alternatively, if an interaction was more significant but occurred sufficiently long ago (either with the satellite or with UGC~5027), the UDG may have already relaxed, and any accreted gas may have had time to settle into a regularly rotating disc (e.g. \citealt{li2026_interactions}). Even without substantial gas accretion, however, tidal perturbations could have driven an internal redistribution of its gas and produced its unusual morphology, although the details of these processes remain uncertain (e.g. \citealt{rory2025,wang2025_interacting}).

Any potential interaction also appears not to have triggered substantial star formation. AGC~198540 has a star formation rate of ${\rm SFR}\,{\approx}\,0.013\,M_\odot\,{\rm yr^{-1}}$ \citep{kado-fong_2022_1}, which is typical for its mass and consistent with the star formation rates of other gas-rich UDGs lying on the star-forming main sequence \citep{kado-fong_2022_1}. However, while this may be at odds with a recent or ongoing interaction, it does not necessarily rule out one in the past, because the timescale traced by the current star formation rate may be much shorter than that of the encounter.

In future work, we will use fly-by simulations to investigate the dynamical evolution of AGC~198540, the fate of its satellite, and determine whether an interaction between AGC~198540 and its satellite or UGC 5027 could reproduce the unusual morphology of the UDG.

\section{Summary and conclusions}
\label{sec:conclusions}
In this work, we introduced the \textit{FRIDA} programme, which comprises high-resolution \hi imaging of 12 isolated gas-rich UDGs, enabled by deep observations carried out with the MeerKAT and VLA interferometers. We presented the galaxy sample that constitutes \textit{FRIDA} and gave an overview of the data reduction process. As expected from their ultra-diffuse nature, \textit{FRIDA} UDGs stand out for their large effective radius and lower average stellar surface densities at their stellar masses (Figure~\ref{fig:reffmstar}).

We measured \hi fluxes from our high-quality data, and find that \textit{FRIDA} covers nearly a factor 20 in \hi mass, from $2\,{\times}\,10^8\,M_\odot$ to $3.6\,{\times}\,10^9\,M_\odot$. We built \hi column-density maps (Figure~\ref{fig:nhimaps}). While \textsc{FRIDA} UDGs are somewhat patchy, they are fairly ordered and symmetric. From the column-density maps, we derived $\Sigma_{\rm HI}(R)$, the azimuthal \hi radial profiles (Figure~\ref{fig:sigma_profs}), which, depending on the UDG, extend out to $7{-}30\,\rm{kpc}$. The $\Sigma_{\rm HI}(R)$ profiles of most \textit{FRIDA} UDGs have a characteristic shape with a central plateau or shallow inner part that reaches $\Sigma_{\rm HI}(R{\rightarrow}0)\,{\approx}\,4\,{\pm}\,1\, M_\odot\,\rm{pc^{-2}}$, which then exponentially declines in the outer parts. The exception is AGC~198540, which also saturates, but has a steeper inner part that then flattens and declines only beyond ${\sim20}\,\rm{kpc}$. 

Using a mock synthetic population, we demonstrated that observed gas-rich UDGs are not preferentially face-on discs despite their selection in surface brightness, and instead should have inclination angles clustering around $45^\circ$ (Figure~\ref{fig:inclinations}). By forward-modelling the \hi morphology of the UDGs in \textit{FRIDA}, we showed that the observed inclination distribution is fully consistent with the theoretical expectations.

We placed the UDGs in two literature \hi mass-size relations, one using $R_{\rm HI}$ (the radius at which $\Sigma_{\rm HI}\,{=}\,1\,M_\odot\,\rm{pc^{-2}}$), and another using $R_{\rm 50,HI}$ (the radius enclosing $50\%$ of $M_{\rm HI}$). \textit{FRIDA} UDGs systematically lie above the best-fitting relations (Figure~\ref{fig:mass_size}). This implies that while UDGs are not extremely different from other galaxies in the \hi mass-size planes, their gas discs tend to be more extended than the average and to have lower average \hi surface densities by $25{-}35\%$.

We analysed a comparison sample of gas-rich galaxies, finding clear progressive trends connecting the \hi mass distribution in galaxies with their offset from the $M_{\rm HI}{-}R_{\rm HI}$ relation and their morphology (Figure~\ref{fig:cumulative}). As galaxies move from the lower to the upper envelope of the relation, their \hi distribution is more spread out, and their surface densities are lower, as expected. Our UDGs are in an intermediate region between the average population and the galaxies with the largest \hi sizes for their \hi masses. Moreover, in terms of their \hi distribution, UDGs follow intermediate trends between late- and early-type discs, showing a large overlap with Sa galaxies, although UDGs have slightly lower central \hi densities and potentially a steeper truncation in the outskirts. \textit{FRIDA} UDGs have a more extended \hi distribution relative to the average population, but they do not appear to have built up an overly extended outer disc, down to our sensitivity.

Finally, we highlighted the case of AGC~198540, a UDG with a markedly different \hi distribution compared to the sample. Our deep \hi data allowed us to discover an unknown companion to the UDG, making this the first time a satellite has been found around a gas-rich UDG. This companion could be responsible for the unusual \hi distribution. We will consider this in future work using theoretical models. \smallskip

\noindent
Overall, \textit{FRIDA} represents a significant leap forward in our understanding of gas-rich UDGs and will enable key kinematic and dynamical analyses that will allow for novel tests of galaxy formation and dark matter models. As our various ongoing projects progress, different data products will become available on \textit{FRIDA}'s website. Along with this paper, we make our \hi radial surface density profiles available.\\

\begin{acknowledgements}

We thank the referee for their careful reading and suggestions. We want to thank the following people: Jing Wang for sharing the data used in \cite{wang16} and for stimulating comments on our results; Yong Shi for sharing the cube of AGC~242019 from \cite{shi2021} and allowing us to include it in our sample; Kyle Oman and Mike Jones for helpful discussions on selection effects; Anna Wright and Arianna Di Cintio for interesting conversations on the \hi properties of simulated UDGs; Rory Smith for insightful comments on galaxy interactions.

PEMP acknowledges funding by the Dutch Research Council (NWO) through the Veni Fellowship VI.Veni.222.364. PEMP is funded by the Research Foundation – Flanders (FWO) through the Fellowship 1219527N.

JIR acknowledges support from STFC grants ST/Y002857/1 and ST/Y002865/1.
This work received funding from the European Research Council (ERC) under the Horizon Europe research and innovation programme (Acronym: FLOWS, Grant number: 101096087).

PEMP would like to thank Stella and Frida S. for the inspiration, love, and companionship.

This paper is based on MeerKAT and VLA data. The MeerKAT telescope is operated by the South African Radio Astronomy Observatory, which is a facility of the National Research Foundation, an agency of the Department of Science and Innovation. VLA is operated by the National Radio Astronomy Observatory. The National Radio Astronomy Observatory and Green Bank Observatory are facilities of the U.S. National Science Foundation operated under cooperative agreement by Associated Universities, Inc. 

We have relied on \cite{bookFilippo}, \cite{binney}, and \cite{graham_driver} for general consultations.
We have used SIMBAD \citep{simbad}, ALADIN \citep{aladin}, NED, and ADS, as well as the tool TOPCAT \citep{topcat} and the Python packages NumPy \citep{numpy}, Matplotlib \citep{matplotlib}, SciPy \citep{scipy}, spectral\_cube \citep{spectral_cube}, cmcrameri \citep{crameri}, and Astropy \citep{astropy}, for which we are thankful.
\end{acknowledgements}

   \bibliographystyle{aa.bst}
   \bibliography{references}

\begin{appendix}
\onecolumn
\section{The isolation of \textit{FRIDA} UDGs}
\label{app:neighbours}

Since our UDGs are gas-rich, they tend to inhabit low-density environments. This is important not only because it eliminates potential
confusion with tidal dwarfs (e.g. \citealt{gray2023}) or debris, but also to ensure that their \hi properties and dynamical state are not heavily affected by recent or ongoing interactions with massive companions. Ten of the UDGs in \textit{FRIDA} come from the catalogue by \cite{leisman2017}, and two more (AGC~198477 and AGC~219199) from \cite{janowiecki2019}. To ensure an isolated sample, \cite{leisman2017} required that their UDGs must not have any neighbour with redshift within $\pm500\,\rm{km\,s^{-1}}$ if their projected separation was not at least $350\,\rm{kpc}$. Instead, \cite{janowiecki2019} did not apply any isolation condition.

To confirm the relative isolation of our UDGs, we query for galaxies in the SIMBAD database \citep{simbad} with a confirmed $cz$ within $\pm500\,\rm{km\,s^{-1}}$ of our sources, and inside a projected radius of $1~\rm{Mpc}$. Three UDGs (AGC~114905, AGC~121790, and AGC~258576) have no neighbours under these conditions, highlighting their isolation. The query for the other galaxies returns between one and a handful of galaxies, and in Table~\ref{tab:neighbours} we list their nearest galaxies in terms of projected spatial and redshift separations, together with their total apparent magnitudes in the $K$ or $B$ band. These nearest galaxies are not necessarily close companions to our UDGs, as even those close in redshift have quite large projected separations (which is strictly a lower limit, since it does not account for the difference in $cz$). The mean (median) distance to the closest projected neighbour is $\sim 434\ (451)\,\rm{kpc}$, and except for the nearest galaxy to AGC~189096\footnote{GAMA 209081 is a small optically blue dwarf, for which the GAMA survey DR4 estimates $M_\ast\sim4\times10^7\, M_\odot$ \citep{driver2022}. The source is not detected in \hi in ALFALFA or in our cubes, from which we estimate a $3\,\sigma$ upper limit $M_{\rm HI}<5\times10^6\, M_\odot$, for linewidths around $20{-}80\,\rm{km\,s^{-1}}$. The galaxy does not appear to have any association with AGC~189096.}, no galaxy closer to ${\sim}280\,\rm{kpc}$ in projection from our UDGs is found. It is clear that our UDGs are isolated from massive galaxies and, down to the detection limits, from other dwarf galaxies as well (but see the main text for the case of AGC~198540).

\begin{table*}[h]
\centering
\caption{Nearest galaxies to \textit{FRIDA} UDGs within 1 Mpc and $500\,\rm{km\,s^{-1}}$.}
\label{tab:neighbours}
\begin{tabular}{l l c c c c}
\hline \noalign{\vskip 1.1pt} 
UDG & Nearest neighbours & Projected sep. & Projected sep. & Spectral sep. & $B$ or $K$ \\ \noalign{\vskip 2.4pt}
 &  & [arcmin] & [kpc] & [$\mathrm{km\,s^{-1}}$] & [mag] \\ \noalign{\vskip 2.4pt}
(1)  & (2) & (3) & (4) & (5) & (6)\\ \noalign{\vskip 2.4pt}
   \hline \noalign{\vskip 1.5pt}  
\noalign{\vskip 1.3pt}
AGC 103435 &  &  &  &  \\ \noalign{\vskip 2.4pt}
 & AGC 103418 & 72.49 & 613.88 & 387.4 & $K:$\,14.63$\pm$0.10 \\ \noalign{\vskip 2.4pt}

AGC 189096 &  &  &  &  \\ \noalign{\vskip 2.4pt}
 & GAMA 209081             & 6.58  & 106.74 & 3.0 & $K:$\,18.07$\pm$0.05  \\ \noalign{\vskip 2.4pt}

AGC 198477 &  &  &  &  \\ \noalign{\vskip 2.4pt}
 & AGC 193218           & 32.01 & 691.98 & -256.1 & $K:$\,15.75$\pm$0.05\\ \noalign{\vskip 2.3pt}
 & AGC 193048 & 43.28 & 935.56 & -23.4 &$B:$\,18.02$\pm$0.36  \\ \noalign{\vskip 2.4pt}

AGC 198540 &  &  &  &  \\ \noalign{\vskip 2.4pt}
 & UGC 5027& 19.09 & 330.65 & -83.7 & $K:$\,12.88$\pm$0.13 \\ \noalign{\vskip 2.4pt}

AGC 219199 &  &  &  &  \\ \noalign{\vskip 2.4pt}
 & UGC 6317                 & 26.22 & 279.71 & -256.0 &$B:$\,15.84$\pm$0.47 \\ \noalign{\vskip 2.3pt}
 & NGC 3633                 & 76.09 & 811.71 & 28.1  & $K:9.88\pm0.03$  \\ \noalign{\vskip 2.4pt}

AGC 226721 &  &  &  &  \\ \noalign{\vskip 2.4pt}
 & UGC 8066                 & 37.30 & 450.71 & -160.0 &$B:$\,15.57$\pm$0.08 \\ \noalign{\vskip 2.3pt}
 & LEDA 214039              & 80.58 & 973.56 & -1.5 & $B:$\,17.01$\pm$0.02  \\ \noalign{\vskip 2.4pt}

AGC 238961 &  &  &  &  \\ \noalign{\vskip 2.4pt}
 & LEDA 1199193             & 23.60 & 451.17 & 294.3 & $K:$\,14.40$\pm$0.77 \\ \noalign{\vskip 2.3pt}
 & AGC 233511 & 43.71 & 835.71 & -35.1 & $K:$\,18.69$\pm$1.06 \\ \noalign{\vskip 2.4pt}

AGC 238984 &  &  &  &  \\ \noalign{\vskip 2.4pt}
 & AGC 238985 & 16.56 & 486.39 & 111.0 & $B:$\,18.12$\pm$0.50\\ \noalign{\vskip 2.3pt}
 & SDSS J135505.09+031745.8 & 16.72 & 490.98 & -11.1 & $K:$\,14.68$\pm$0.36 \\ \noalign{\vskip 2.4pt}

AGC 242019 &  &  &  &  \\ \noalign{\vskip 2.4pt}
 & AGC 249449              & 65.16  & 493.23 & -127.7 & $B:$\,18.25$\pm$0.50 \\ \noalign{\vskip 2.4pt}
 & LEDA 1206999             & 88.49  & 669.84 & -6.4 & $B:$\,18.03$\pm$0.50  \\ \noalign{\vskip 2.4pt}
   \hline
    \end{tabular} 
    \tablefoot{
    (1): \textit{FRIDA} UDG. 
    (2): Nearest neighbour in projected distance and in redshift separation.
    (3): Projected separation between the UDG and its neighbour, as given by SIMBAD coordinates.
    (4): Same as (3), but converted to kpc assuming the distance to the UDG.
    (5): Difference in $cz$ (i.e. in systemic velocities between the UDGs and the nearest neighbours), as given by SIMBAD. 
    (6): Total apparent magnitude in the $K$ or $B$ bands, as quoted in the HyperLeda database \citep{hyperleda}.
    The galaxies AGC~114905, AGC~121790, and AGC~258976 do not have confirmed neighbours within our search limits.
    }
\end{table*}

\section{Optical images}
\label{app:optical}
\begin{figure}[h]
    \centering
    \includegraphics[width=1\linewidth]{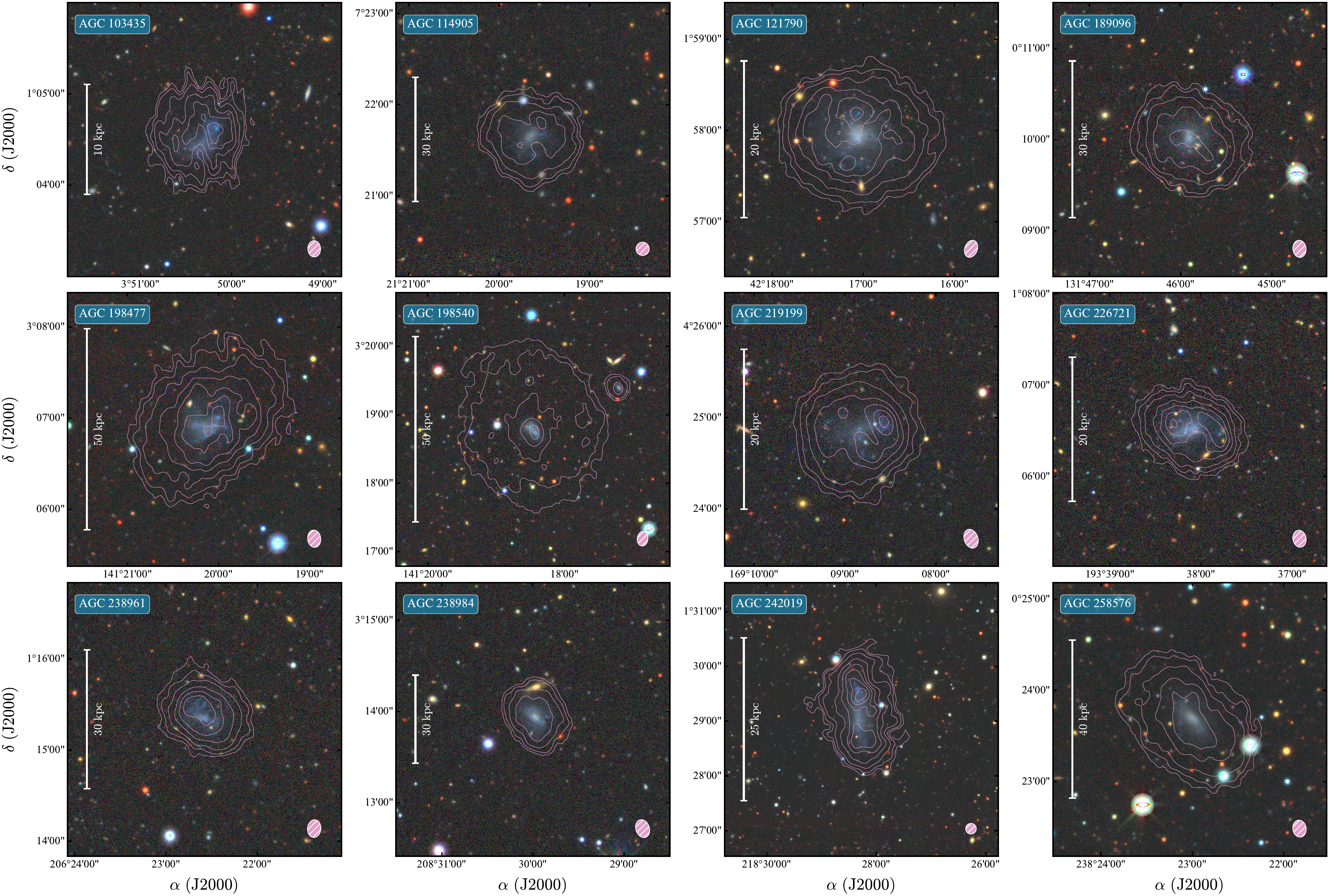}
    \caption{\hi contours (pink, beam at the bottom right) overlaid on DECaLs optical RGB stamps for the UDGs in \textit{FRIDA}. The white scales represent various sizes comparable to the \hi extent.}
    \label{fig:stamps}
\end{figure}

Figure~\ref{fig:stamps} shows stamps of the UDGs in \textit{FRIDA}. In particular, we show \hi column-density contours (Section~\ref{sec:HIprop}) on top of DECaLs RGB optical images. As part of \textit{FRIDA}, we have undertaken an observational campaign with the GTC to obtain ultra-deep imaging of the UDGs' stellar discs, as done in the past with AGC~114905 \citep{agc114905_deep}; about three-quarters of the sample has been observed, and the data will be presented in an upcoming paper. This ultra-deep imaging is crucial, as \cite{agc114905_deep} demonstrated that DECaLs' depth is not always sufficient to trace the extended underlying stellar disc of UDGs. 

\section{Recovery of \hi fluxes compared to single-dish data}
\label{app:fluxes}
\begin{figure}[h]
    \begin{minipage}{0.46\linewidth}
        \includegraphics[width=\linewidth]{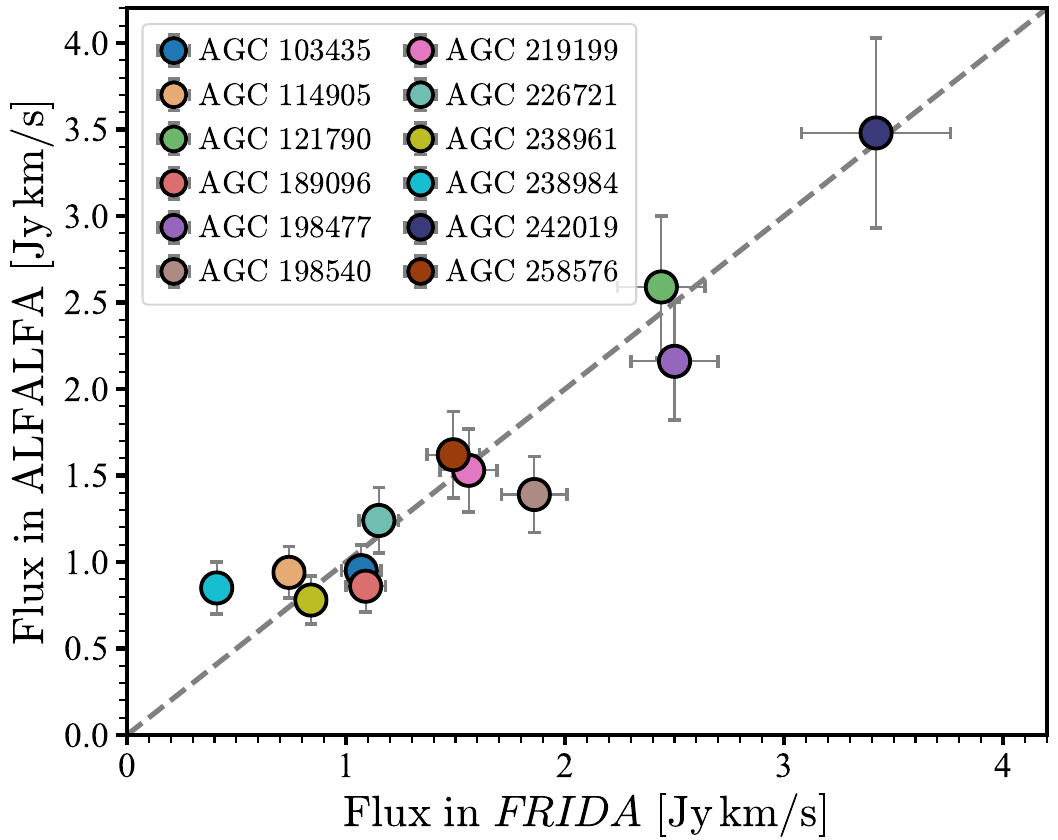}
    \end{minipage}
    \hspace{0.02\linewidth}
    \begin{minipage}{0.52\linewidth}
        \captionof{figure}{Comparison between the fluxes measured in this work and those from ALFALFA data measured by \cite{Yu_Ho_2022}. The grey dashed line shows the 1:1 relation.
        }
        \label{fig:flux_compar}
    \end{minipage}
\end{figure}

In Figure~\ref{fig:flux_compar} we compare the \hi fluxes measured from our MeerKAT and VLA data, and those recovered from ALFALFA single-dish data by \cite{Yu_Ho_2022}. There is good agreement between the two sets of measurements, and the interferometric observations do not systematically miss flux, as expected given MeerKAT's good coverage of short baselines and our multi-configuration VLA setup. The median and mean ratio between the two flux measurements is $1.0$, with a standard deviation of $22\%$. The standard deviation is mainly driven by AGC 198540, for which we recover $37\%$ more flux, and by AGC~238984, for which our flux is $48\%$ lower. Closer inspection of the galaxies suggests that noisy ALFALFA spectra drive these differences; the ALFALFA measurements become consistent with ours within ${\sim}\,2\,{-}\,2.5\,\sigma$.

\section{\hi inclination angles in \textsc{FRIDA} }
\label{app:bias_inclination}
\begin{figure}
    \begin{minipage}{0.75\linewidth}
        \includegraphics[width=\linewidth]{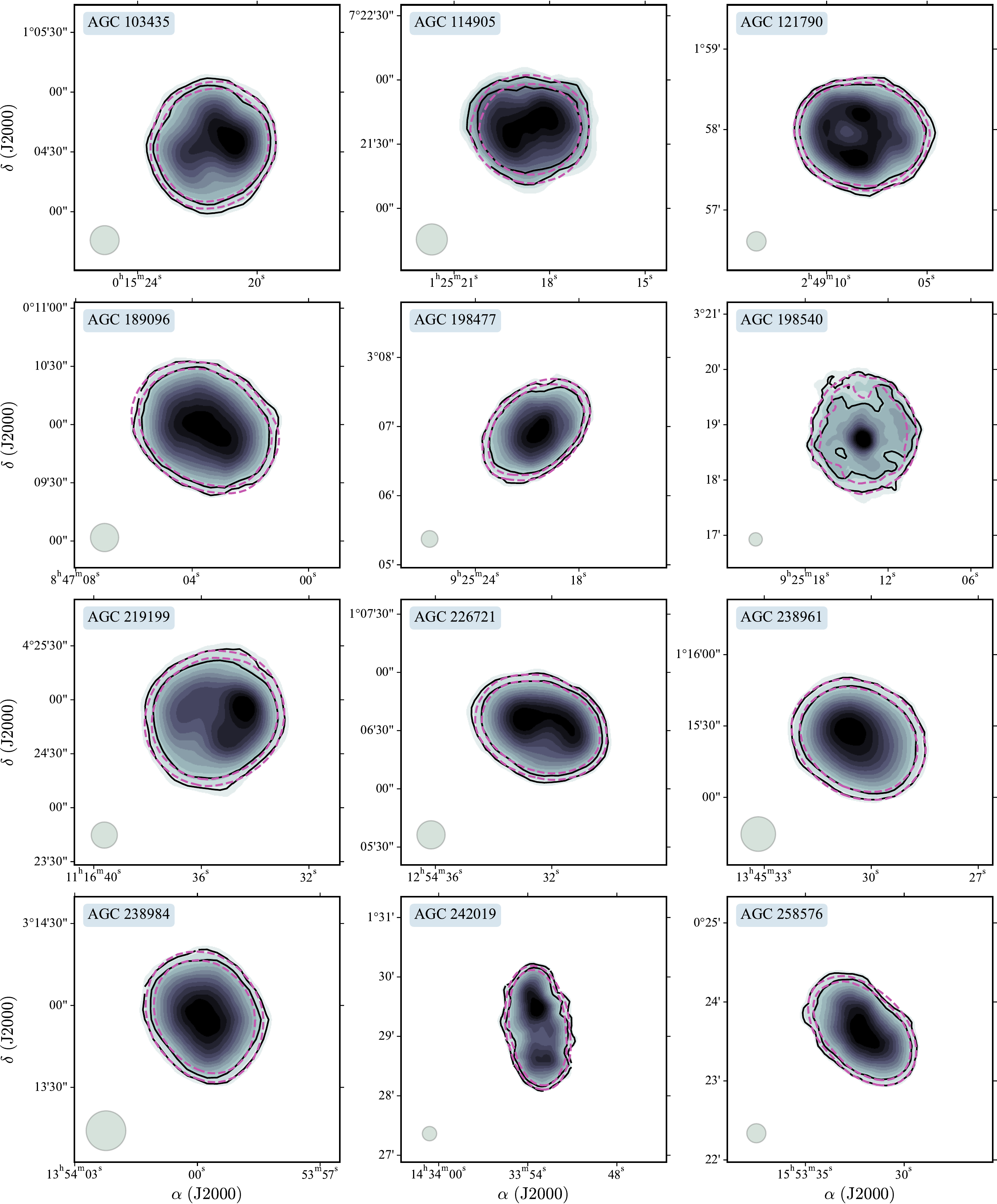}
    \end{minipage}
    \hspace{0.03\linewidth}
    \begin{minipage}{0.2\linewidth}
        \captionof{figure}{Comparison between the moment-0 maps (background shades and black contours) used to measure the geometrical parameters of the UDGs in \textit{FRIDA} and the best-fitting \texttt{cannubi} models (purple contours).}
        \label{fig:cannubi_compar}
    \end{minipage}
\end{figure}

Figure~\ref{fig:cannubi_compar} compares the moment-0 maps (at a resolution of 15 arcsec) with the best-fitting \texttt{cannubi} models used to derive the geometrical parameters (Section~\ref{sec:painc}). The robustness of our method is illustrated by the good agreement between the data and the model, particularly in the outer parts, where the inclination and position angle are better constrained.

\end{appendix}

\end{document}